\documentclass[12pt]{iopart} % IOPScience class

\usepackage{graphicx}
\usepackage{siunitx} % For SI units
\usepackage[version=4]{mhchem} % Chemical formulae
\usepackage{booktabs} % Table \rules
\usepackage{subcaption} % For subfigures
\usepackage[width=.9\textwidth]{caption} % For captioning figures
\usepackage{hyperref} % Referencing
\usepackage[section]{placeins} % Figure placing
\usepackage{iopams} % Recover \mathbb forbidden because of AMS conflicts
\usepackage{adjustbox}
\usepackage{doi}
\usepackage{orcidlink} % For orcid links
\usepackage{cprotect} % For defining verbs inside captions of figures/tables
\hypersetup{colorlinks,
    citecolor=blue,
    filecolor=blue,
    linkcolor=blue,
    urlcolor=black
}

\newcommand{\tss}[1]{\textsuperscript{#1}} % \textsuperscript in fewer chars
\def\L{\mathcal{L}} % Ease of equations
\def\S{\mathcal{S}} % Ease of equations
\def\Zeff{Z_\textrm{eff}} % To save space for effective Z
\def\mvm{\mu_\textrm{VM}} % Save space for VM labels
\def\mueff{\langle\mu\rangle} % Effective attenuation

\usepackage{subfiles} % Always load last to avoid problems

\begin{document}

\title[]{Proton stopping power images from Monte Carlo simulated dual-energy CT scans}

\author{Adam Zieser \orcidlink{0000-0001-7618-3382}\tss{1}, Ugur Akgun \orcidlink{0000-0002-9850-4164}\tss{1,2}, Yasar Onel \orcidlink{0000-0002-8141-7769}\tss{1}}

\address{\tss{1} Department of Physics \& Astronomy, University of Iowa, Iowa City, USA \\
\tss{2} Physics Department, Coe College, Cedar Rapids, USA}

\ead{adam-zieser@uiowa.edu}

\vspace{10pt}

\begin{abstract}
We test the feasibility of calculating proton stopping power (SP) from a set of virtual monochromatic (VM) CT images, created with Monte Carlo-generated CT scans, using a dual-energy CT-to-SP procedure. The Monte Carlo CT simulations, with \SI{70}{\kilo\volt p} and \SI{150}{\kilo\volt p} spectra, were modeled on the x-ray tube parameters of the SOMATOM Force dual-source DECT scanner and the Gammex RMI 467 tissue calibration phantom. Reconstructed SP images were created with two VM x-ray images, and the monochromatic energy pairs were chosen to minimize the SP residual errors using the root-mean-squared-error (RMSE). The contrast of phantom inserts has also been examined, in addition to relative errors of the average reconstructed insert values (which are calculated using the known electron densities and chemical compositions of each material, provided by the manufacturer). Results were also compared to a similar dual-energy-based SP conversion procedure for comparison. Modest reductions in errors and noise were seen in almost all inserts in both low- and high-density configurations of the phantom, with a visible reduction in beam-hardening streaking artifacts. Our Monte Carlo simulated DECT scans confirm the feasibility of previously proposed methods of DECT-based $ S(\rho_e, Z) $-determination. By utilizing VM images and only optimizing the RMSE over specific inserts, the residual error across the most vital bodily tissues may be reduced when extremely low- or high-density tissues are known to be present.
\end{abstract}
%\vspace{2pc}
%\noindent{\it Keywords\/}: dual-energy CT, monte carlo, stopping power, virtual monochromatic

%\submitto{Biomed. Phys. Eng. Express}
%\submitto{\PMB}

\section{Introduction}

The depth-dose profile of MeV-scale protons in matter---which are characterized by a low-dose sub-peak region followed by a sharp Bragg peak and distal falloff---allow for the deposition of high radiation doses to tumors while sparing healthy tissue, making protons particularly attractive for use in radiotherapy. \cite{Smith2006,Newhauser2015} To utilize this form of treatment, accurate proton stopping power (SP) images of a patient are required, which are typically acquired through use of a CT Hounsfield Unit (HU) to stopping power calibration curve, a process referred to as the ``stoichiometric method." \cite{Schneider1996,Schaffner1998} This calibration procedure introduces problematic SP range errors on the order of \SI{2.5}{\percent} \cite{Meinsdottir2019} to \SI{3.5}{\percent}. \cite{Moyers2001, Moyers2010, Paganetti2012, Yang2012} Absolute range errors are frequently assumed in addition to these relative errors, with values anywhere from \SI{1}{\mm} to \SI{7}{\mm}. \cite{Meinsdottir2019, Taasti2019} The calibration curve fitting procedure is also unique to each scanner's x-ray tube spectrum, and is a somewhat-arbitrary process, as a curve must be chosen to prioritize certain tissues due to the degeneracy of the $ \mathrm{SP}(\mathrm{HU}) $ function: the same HU value may correspond to multiple SP values. As such, SP may benefit from being parameterized by more than one quantity. \cite{Yang2010,Yang2012,Li2017} Stoichiometric-derived SP images are also subject to the same sources of error inherent to conventional x-ray CT images, including beam hardening streaking artifacts along the beam paths that contain multiple high-density objects. Recently, dual-energy CT (DECT) has been explored as a  possible avenue of improvement on the stoichiometric calibration in determining patient SP. \cite{Bar2017}

DECT utilizes dual-layer detectors or multiple incident x-ray spectra to extract complete knowledge of an object's attenuation coefficient at any (and all) medically relevant energies. \cite{vanElmpt2016, Johnson2012, Rajiah2020} With this information, it is also possible to reconstruct electron density (ED) and effective atomic number (EAN), which together provide sufficient information to reconstruct patient SP images. \cite{Rutherford1976, Heismann2003} Many commercial DECT scanners are now equipped to reconstruct ED and EAN natively, and SPs derived from them have been investigated for use in proton therapy. \cite{Taasti2018, Mohler2018} Different variations of this ``$ \rho_e Z $-decomposition" procedure have been proposed, with some requiring calibration based on each commercial scanner used, \cite{Saito2012, Landry2011_1, Landry2011_2, Hunemohr2014_1, Hunemohr2014_2, Bourque2014,  Han2016, Lalonde2016, Taasti2016} and with some requiring only x-ray tube spectral information. \cite{Heismann2003, Heismann2009, VanAbbema2015} A number of models are based on a parameterization of the x-ray attenuation coefficient by Jackson and Hawkes, \cite{JacksonHawkes1981} such as that of Torikoshi et al., \cite{Torikoshi2003} Bazalova et al., \cite{Bazalova2008} or Yang et al. \cite{Yang2010} While spectrum-utilizing methods are typically less accurate than calibration-based ones, \cite{Bar2017} they are particularly well-suited for Monte Carlo simulations of CT acquisitions, as the incident x-ray spectrum is the only scanner-specific parameter that needs to be modeled. 

Noteworthy is the fact that these spectra-dependent, DECT-derived SP images eliminate the functional degeneracy of the stoichiometric calibration, but may or may not inherently account for beam-hardening error, as the decomposition methods often assume identical entrance and exit spectra and do not account for the effect theoretically, \cite{Mahnken2009, Yang2010, Landry2011_2} instead opting for spectral pre-filtration. \cite{Bazalova2008} or other empirical methods. Alternatively, if a pair of sufficiently different x-ray projections are first used to create a pair of virtual monochromatic (VM) attenuation images using established DECT techniques, \cite{Alvarez1976, Macovski1978} then beam-hardening errors can be removed prior to calculation of ED and EAN, without the need for additional corrections or filtration. Creation of VM images can even be carried out without knowledge of tube spectra---provided that certain calibration measurements are carried out. \cite{Stenner2007, Hao2012} Recently, N{\"a}smark and Andersson \cite{Nasmark2021} were the first to apply the Jackson and Hawkes attenuation parameterization to DECT-generated VM images, acquiring low- and high-\SI{}{\kilo\volt p} images on a rapid \si{\kilo\volt}-switching GE Revolution CT scanner (GE Healthcare, Waukesha, WI, USA). We present a slight variation on this method here, further confirming its feasibility.
%
% -------------------------- Figure --------------------------------
\begin{figure}[t]
\centering
\includegraphics[width=0.7\textwidth]{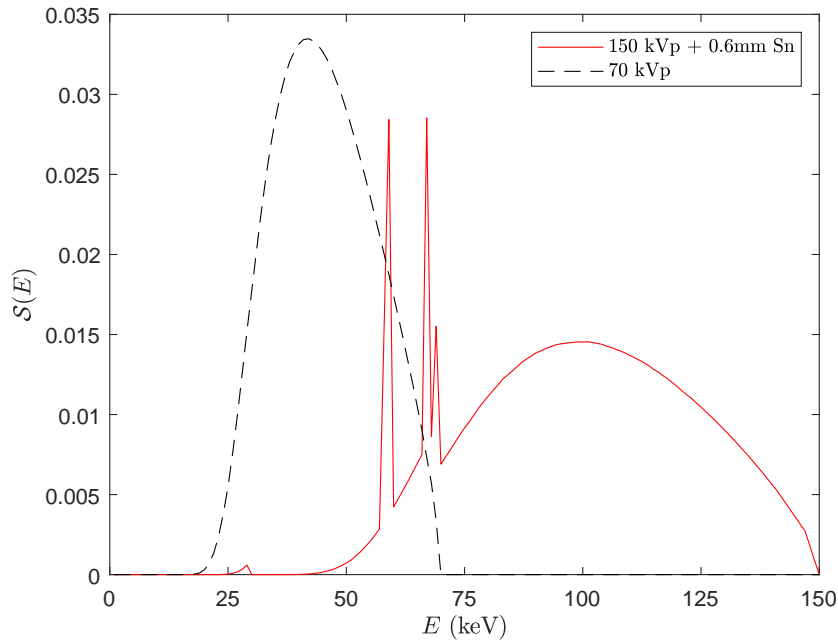}
\caption{The \SIlist{70;150}{\kilo\volt p} spectra used in this study, generated using SpekCalc, \cite{SpekCalc1, SpekCalc2, SpekCalc3} with \SI{1}{\keV} bin size. \label{fig:Spectra}}
\end{figure}
% ------------------------------------------------------------------
%

This variation on N{\"a}smark and Andersson's procedure combines multiple DECT and $ \rho_e Z $-decomposition models and uses the root-mean-square (RMSE) to create SP images that optimize for specific tissues. To simulate a realistic scenario, Monte Carlo simulations of low- and high-energy CT scans that mimic the spectral properties of the Siemens SOMATOM Force dual-source CT scanner (Siemens Healthineers, Erlangen, Germany) were performed. These data sets were used to construct pairs of VM attenuation images according to the DECT process proposed by Zou and Silver, \cite{Zou2008, Zou2009} which does not require a scanner-specific calibration. ED and EAN images were then created according to the monoenergetic attenuation-decomposition presented by Torikoshi et al. \cite{Torikoshi2003} After converting EAN to mean excitation energy (MEE) via a conversion curve proposed by Bourque et al., \cite{Bourque2014} proton SP images $ S(\rho_e,I) $ were calculated according to the Bethe-Bloch formula. The RMSE aids in choosing the optimal low- and high-energy pairs for the VM attenuation images. The effectiveness of this procedure is tested on two different configurations of the Gammex RMI 467 tissue calibration phantom (Gammex Inc., Middleton, WI): one standard configuration recommended by the manufacturer, and one with a number of water-equivalent inserts replaced with grade 2 titanium alloy to introduce prominent beam hardening artifacts. These setups are hereafter referred to as the \textit{standard} and \textit{titanium} configurations. The accuracy of the reconstructed images have been examined using contrast ratios of each phantom insert against neighboring solid-water background, and residual errors from the ground truth reference stopping powers---which are calculated using EDs and elemental compositions provided by the phantom manufacturer. The results are compared to a similar $ \rho_e Z $-decomposition method which does not first construct VM attenuation images, \cite{Bazalova2008} showing that the procedure may be able to improve upon previous dual-energy $ S(\rho_e, Z) $ conversion methods. The proposed procedure is hereafter referred to as the ``VM-SP" method, with the reference method referred to as the ``polychromatic CT-SP" method.
\section{Materials and Methods}
\subsection{DECT simulation parameters and virtual monochromatic images}

For x-ray energies above bodily tissue K-edges but below the pair-production threshold (\SI{1.022}{\MeV}), the x-ray attenuation coefficient can be approximated \cite{Alvarez1976, Alvarez2019} using the basis-material decomposition 
%
% -------------------------- Equation ------------------------------
\begin{eqnarray}
\mu(\vec{r}, E) = a_1(\vec{r}) \mu_1(E) + a_2(\vec{r}) \mu_2(E), \label{eq:Alvarez}
\end{eqnarray}
% ------------------------------------------------------------------
%
where $ \mu_1 $ and $ \mu_2 $ are the pre-tabulated attenuation coefficients for the two basis-materials which interact primarily as photoelectric absorbers and incoherent scatterers, respectively, and  $ a_1 $ and $ a_2 $ are material-dependent coefficients. If two CT scans of an object are acquired with known high- and low-energy spectra $ \S_i, \ i \in \{ H, L \} $, then path-dependent projection data seen by the detector are given by
%
% -------------------------- Equation ------------------------------
\begin{eqnarray} 
P_i(\L) = - \ln \int \S_i(E) \exp \bigg( - \sum_{m=1}^2 \mu_m(E) A_m(\L) \bigg) \rmd E,
\label{eq:DECTProjection}
\end{eqnarray}
% ------------------------------------------------------------------
%
along ray $ \L $, with basis-material-equivalent thicknesses $ A_m $ containing all position-dependence of the projections:
%
% -------------------------- Equation ------------------------------
\begin{eqnarray}
A_m(\L) = \int_{\L} a_m(\ell) \rmd \ell.
\label{eq:MET}
\end{eqnarray}
% ------------------------------------------------------------------
%
Given that the two spectra are sufficiently non-overlapping and the basis materials are well-chosen (according to the conditions described by Alvarez \cite{Alvarez2019}) the two coupled integral equations \eref{eq:DECTProjection} can be approximately inverted to solve for the equivalent thicknesses, and the two coefficients $ a_1 $ and $ a_2 $ can be found via standard CT image reconstruction techniques by inverting the Radon transform in \eref{eq:MET}. \Eref{eq:MET} is invertible in a tomographic slice if $ A_m $ is known for all possible rays $ \L $ in the space of straight lines in $ \mathbb{R}^2 $. This is known as DECT carried out in the ``imaging domain."

In this work, water was chosen as the scattering basis, and iodine as the photoelectric basis---the latter outperforming calcium, aluminum, and titanium in our tests, despite iodine's high K-edge and atomic number. All basis attenuation data were taken from the NIST XCOM \cite{XCOM} database and sampled at \SI{1}{\keV} intervals. \SIlist{70;150}{\kilo\volt p} spectra were generated with SpekCalc \cite{SpekCalc1, SpekCalc2, SpekCalc3} according to the properties of the Siemens Vectron x-ray tube (pictured in \fref{fig:Spectra}). Both spectra are generated with tungsten anodes, \SI{7}{\degree} anode angle, and \SI{6.8}{\mm} of aluminum-equivalent filtration. The high-energy spectrum was also filtered with an additional \SI{0.6}{\mm} of tin to eliminate low-energy wavelengths. These spectra were sampled in \SI{1}{\keV} energy bins with \num[exponent-product = \times]{1e9} particle histories simulated per projection, in the Monte Carlo radiation transport code MCNP6.1.1. \cite{MCNP6.1, MCNP6.1.1} Photons were transported using an appropriate physics treatment, such that photoelectric absorption, coherent scattering, and incoherent scattering have been modeled, but photonuclear interactions were disabled. It is assumed that electrons  deposit all energy locally in charged particle equilibrium, \cite{Gu2009, DeMarco2005} so only photons were transported. The histories were uniformly sampled from a \SI{14.1}{\cm} mono-directional line source to utilize a more simple reconstruction algorithm. Only one two-dimensional slice of the phantom (\SI{0.5}{mm} thick) has been simulated and reconstructed. As such, all x-ray histories that exited the simulated slice's thickness were terminated.
%
% -------------------------- Figure -------------------------------
\begin{figure}[t]
\centering
\includegraphics[width=0.75\textwidth]{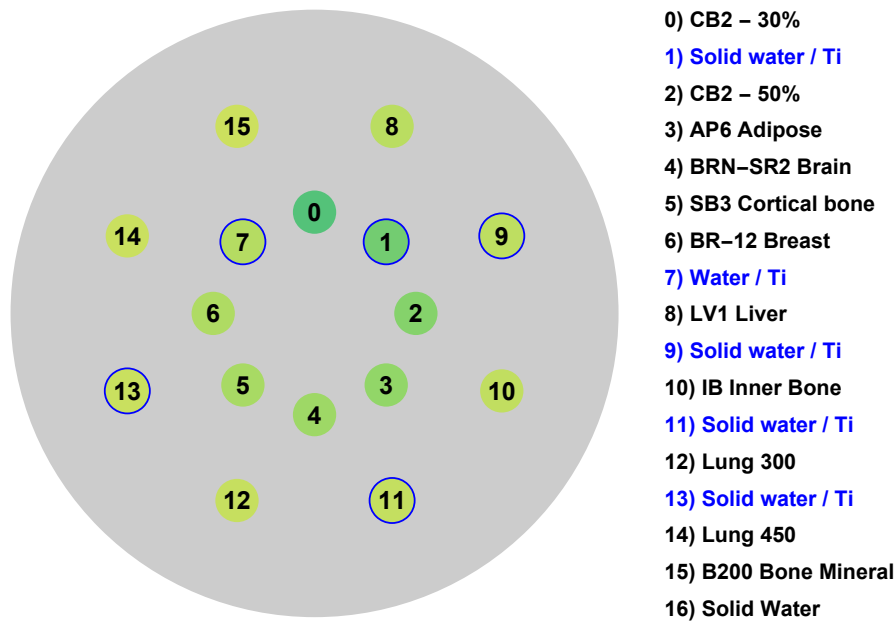}
\caption{Geometry and inserts of one 2D slice of the Gammex RMI 467 tissues calibration phantom in both configurations. Inserts 1, 7, 9, 11, and 13 change according to the insert configuration being used. \label{fig:GammexFig}}
\end{figure}
% ------------------------------------------------------------------
%

For the same reason, the detector was modeled as a \SI{14.1}{\cm} wide, straight line of $ \SI{0.5}{\mm} \times \SI{0.5}{\mm} \times \SI{0.5}{\mm}  $ volume-flux mesh tally voxels using the \verb|FMESH| tally option in MCNP, approximating a straight-line equivalent of one strip of the Siemens Stellar\tss{Infinity} detector---but with perfect detector response and zero quantum noise. Output from SpekCalc spectra combined with acceptable tube current and exposure times indicate that on the order of \num[exponent-product = \times]{1e9} x-ray histories per projection is realistic for a CT scan, and is supported by the convergence of all MCNP mesh tally bins with coefficient of variation well below \num{0.05}. Additionally, preliminary simulations showed that an anti-scatter grid with a 10:1 grid ratio do not significantly alter the results from those performed without such a grid, and was therefore omitted to improve tally convergence and reduce computation time.

The image reconstruction grid was set to \num{281} by \num{281} pixels, matching the dimension of the reconstruction grid to the number of pixels of the detector. The inversion of the equation set \eref{eq:DECTProjection} was performed with a custom script in Matlab (Mathworks, Natick, MA) using an implementation of the perturbative DECT method of Zou and Silver, \cite{Zou2008,Zou2009} with 20 iterations of the algorithm to ensure convergence. The basis coefficients were found via filtered backprojection (FBP) using an edited copy of the Matlab \verb|iradon| function. VM images were then constructed using \eref{eq:Alvarez} for any energy in the diagnostic range. The phantom studied was scaled to \SI{40}{\percent} of its original size, with the inserts scaled to one-third size, and was modeled using two separate insert configurations to study the effects of beam hardening. As seen in \fref{fig:GammexFig}, one is a suggested configuration according to the user manual, and one replaces all water-equivalent inserts with grade \num{2} titanium alloy. The elemental compositions, densities, and electron densities of each material are provided by the manufacturer (see \tref{tab:GammexTable}).

%
% -------------------- Table -----------------------------------
\begin{table}[tb]
\centering
\caption{The physical properties and elemental compositions (by weight) of the inserts of the Gammex RMI 467 phantom. Mass and relative electron densities were taken from the phantom user guide. All compositions (with the exception of titanium alloy) have been taken from the manufacturer, as listed in Van Abbema et al. \cite{VanAbbema2015}
\label{tab:GammexTable}}
\subfile{anc/PhantomTable.tex}
\end{table}
% --------------------------------------------------------------
%
\subsection{$ \rho_e Z $-decomposition} \label{RhoZ}
To generate ED and EAN images using low- and high-energy pairs of VM DECT images, the decomposition of Jackson and Hawks \cite{JacksonHawkes1981} presented by Torikoshi et al. \cite{Torikoshi2003} was used,
%
% -------------------------- Equation ------------------------------
\begin{eqnarray}
\mvm (E) = \rho_e \Big( \Zeff^4 F(\Zeff, E) + G(\Zeff, E) \Big). \label{eq:MuDecomp}
\end{eqnarray}
% ------------------------------------------------------------------
%
Assuming that the photoelectric and scattering terms $ F $ and $ G $ are weakly dependent on $ \Zeff $, the EAN can be approximately solved with a root finder according to
%
% -------------------------- Equation ------------------------------
\begin{eqnarray}
\Zeff^4 - \frac{ \mvm (E_H) G(\Zeff, E_L) - \mvm(E_L) G(\Zeff, E_H) }{ \mvm (E_L) F(\Zeff, E_H) - \mvm (E_H) F(\Zeff, E_L) } = 0, \label{eq:Zeff}
\end{eqnarray}
% ------------------------------------------------------------------
%
which is then re-inserted into the following equation to calculate the ED:
%
% -------------------------- Equation ------------------------------
\begin{eqnarray}
\rho_e = \frac{ \mvm (E_L) F(\Zeff, E_H) - \mvm (E_H) F(\Zeff, E_L) }{ F(\Zeff, E_H) G(\Zeff, E_L) - F(\Zeff, E_L) G(\Zeff, E_H) }. \label{eq:rhoe}
\end{eqnarray}
% ------------------------------------------------------------------
%
NIST XCOM data were used to find the photoelectric and scattering partial interaction mass-attenuations $ \mu \mathbin{/} \rho $ for all elements from $ Z = $ \numrange{1}{23}, from which $ F $ and $ G $ are calculated,
%
% -------------------------- Equation ------------------------------
\numparts
\label{eq:XCOMData}
\begin{eqnarray}
& \frac{F(Z,E)}{N_A} = \frac{A}{Z^5} \frac{\mu_\textrm{Ph}}{\rho}, \label{eq:photoF} \\ 
& \frac{G(Z,E)}{N_A} = \frac{A}{Z} \bigg( \frac{\mu_\textrm{coh}}{\rho} + \frac{\mu_\textrm{inc}}{\rho} \bigg). \label{eq:scatterG}
\end{eqnarray}
\endnumparts
% ------------------------------------------------------------------
%
%
% -------------------------- Figure -------------------------------
\begin{figure}[t]
\centering
\includegraphics[width=0.65\textwidth]{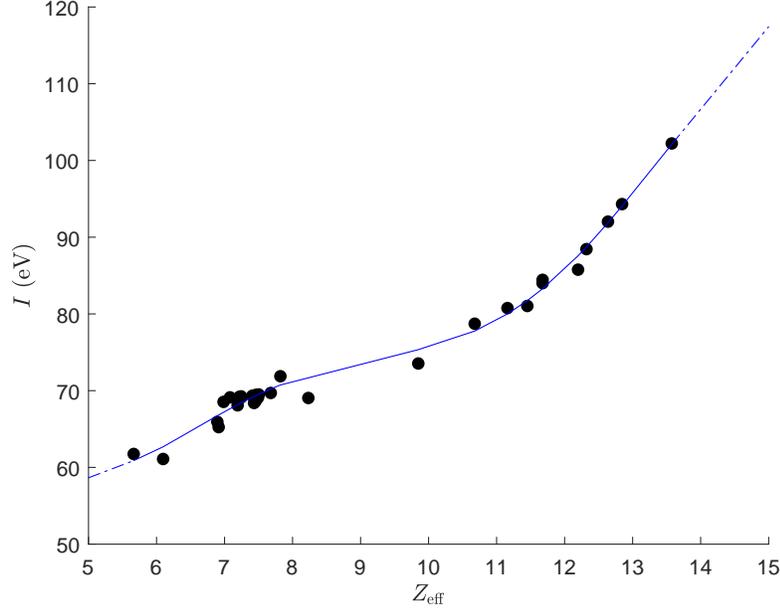}
\caption{The empirical 5th-order polynomial fit of MEE versus EAN according to the theory of Bourque et al. \cite{Bourque2014} using tissue data from White et al., \cite{ICRU44, White1987} with thyroid tissue excluded due to the presence of iodine. The bounds $ Z_0 $ and $ Z_\mathrm{end} $ in \eqref{eq:MEE} are approximately \num{5.66} and \num{11.16}, respectively.
\label{fig:MEECurve}}
\end{figure}
% ------------------------------------------------------------------
%
In order to derive MEE for use in SP calculation, Yang et al. \cite{Yang2010} and Bourque et al. \cite{Bourque2014} have established an empirical polynomial dependence on EAN. The 5th-order polynomial fit of Bourque was used in this study, as the $ I(\Zeff) $ curve more smoothly fit tissues in our reconstructions with intermediate-value EAN (around \numrange{8}{10}). For a data set used in fitting, $ \{ (Z_{0},I_{0}), \, \dotsc , \, (Z_{\textrm{end}},I_{\textrm{end}}) \} $,
%
% -------------------------- Equation ------------------------------
\begin{eqnarray}
\fl I(\Zeff) = 
\begin{cases}
c_1 + c_2 \Zeff & \Zeff < Z_0 \\
c_3 + c_4 \Zeff^1 + c_5 \Zeff^2 + c_6 \Zeff^3 + c_7 \Zeff^4 + c_8 \Zeff^5 & Z_0 \leq \Zeff \leq Z_{\textrm{end}} \\
c_9 + c_{10} \Zeff & Z_{\textrm{end}} < \Zeff
\end{cases}.
\label{eq:MEE}
\end{eqnarray}
% ------------------------------------------------------------------
%
To determine the fit coefficients, tissue data from White et al. \cite{ICRU44, White1987} was used to calculate attenuation coefficients from XCOM according to their elemental compositions, which were then used in \eref{eq:Zeff}. Using the EAN values from this set of tissues, the fit coefficients were found to be: $ c_1 = \num{41.8369}, \ c_2 = \num{3.3621}, \ c_3 = \num{586.5268}, \ c_4 = \num{-326.7633}, c_5 = \num{77.0502}, \ c_6 = \num{-8.6343}, \ c_7 = \num{0.4646}, \ c_8 = \num{-0.0096}, \ c_9 = \num{-43.9449}, \ c_{10} = \num{10.7587} $. All relevant quantities were calculated via Matlab script, with \eref{eq:Zeff} solved using the built-in \verb|fsolve| function using the Levenberg–Marquardt algorithm.
\subsection{Proton stopping powers and analysis}

\begin{figure}[t]
\begin{center}
\includegraphics[width=0.65\textwidth]{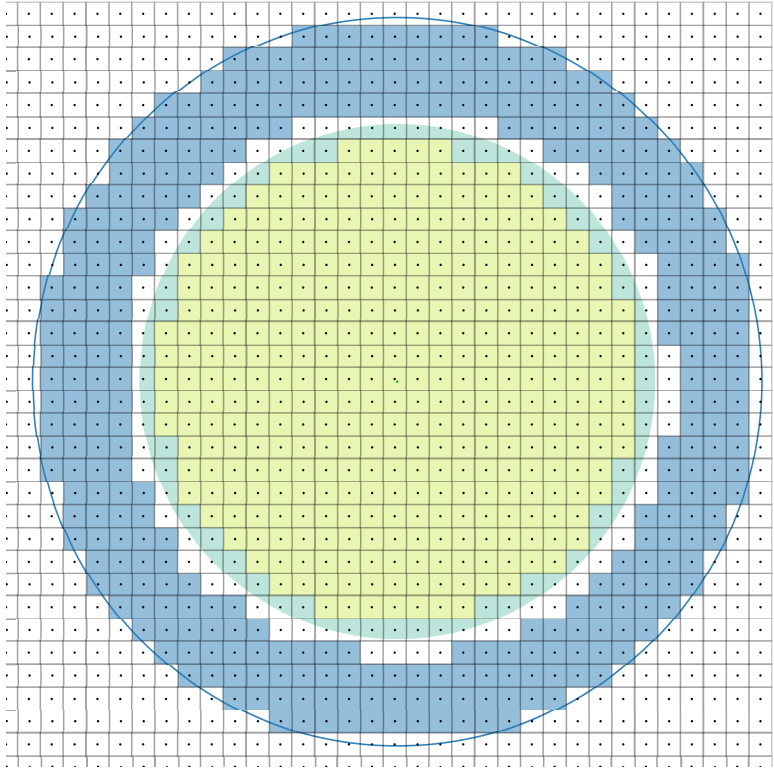}
\caption{An example diagram of all pixels that are completely contained within a given circular insert (in yellow). The surrounding pixels are an equal-area number of pixels of the bordering solid water region for use in contrast calculation (in blue). Pixel centers and the insert boundary (in green) are shown for visual clarity.
\label{fig:PixelGeometry}}
\end{center}
\end{figure}

Proton SPs were calculated with reconstructed ED and MEE values using the Bethe-Bloch formula, \cite{Newhauser2015}
%
% -------------------------- Equation ------------------------------
\begin{eqnarray}
S(\rho_e, I, E) = 4 \pi r_e^2 m_e c^2 \frac{\rho_e}{\beta(E)^2} \bigg[ \ln \bigg( \frac{2 m_e c^2}{I} \frac{\beta(E)^2}{1-\beta(E)^2} \bigg) - \beta(E)^2 \bigg].
\label{eq:Bethe}
\end{eqnarray}
% ------------------------------------------------------------------
%
To calculate the assumed ground truth from which the residual errors are calculated, \eref{eq:Bethe} was used with the tissue EDs provided by the phantom user's guide and MEEs calculated by elemental composition via Bragg additivity formula. With fractions by weight $ \lambda_m $, this is given by:
%
% -------------------------- Equation ------------------------------
\begin{eqnarray}
\ln I = \Bigg( \sum_{m} \lambda_m \frac{Z_m}{A_m} \ln I_m \Bigg) \mathbin{\Bigg/} \Bigg( \sum_m \lambda_m \frac{Z_m}{A_m} \Bigg). \label{eq:BraggI}
\end{eqnarray}
% ------------------------------------------------------------------
%
Elemental MEE values were taken from Seltzer and Berger \cite{SeltzerBerger1981} (chosen to be consistent with XCOM data). To evaluate the accuracy of each reconstruction, the average SP value of the pixels in each insert $ i $, $ \bar{S}_i $, were compared to their corresponding theoretical values, $ S_\mathrm{Ref,i} $, by their residual errors:
%
% -------------------------- Equation ------------------------------
\begin{eqnarray}
\mathrm{RE}_i = \frac{ \bar{S}_i - S_\mathrm{Ref,i} }{ S_\mathrm{Ref,i} } \times \SI{100}{\percent}. \label{eq:delta}
\end{eqnarray}
% ------------------------------------------------------------------
%
These averages are found by summing over all pixels known to be inside an insert, found by iterating over all nearby pixels (see \fref{fig:PixelGeometry}) and counting those which lie entirely within the insert. For an insert $ i $ with a radius $ r_i $ and center coordinates $ (y_i,z_i) $, a pixel with center $ (y,z) $ is considered contained within the insert if the following condition is met:
%
% -------------------- Pixel boundary --------------------------
\begin{eqnarray}
(y - y_i)^2 + (z - z_i) < (r_i - \sqrt{2} / 2)^2.
\label{eq:InPixel}
\end{eqnarray}
% --------------------------------------------------------------
%
For each insert, this is satisfied by approximately \num{350} pixels. Standard deviations are also calculated to estimate confidence intervals of one standard error of the mean.
%
% -------------------------- Figure -------------------------------
% ------------------------------------------------------------------
%
Using the same information, contrast ratios were found for each insert (with the background being the surrounding solid water pixels denoted by $ B $) using the following definition:
%
% -------------------------- Equation ------------------------------
\begin{eqnarray}
C_i = \frac{| \bar{S}_i - \bar{S}_{B,i} |}{\sqrt{ \sigma_i^2 + \sigma_{B,i}^2 }},
\label{eq:CNR}
\end{eqnarray}
% ------------------------------------------------------------------
%
with $\sigma_i $ being the standard deviation of insert pixels, and $ \sigma_{B,i} $ being the water background/border standard deviation. A punctured disk of water pixels surrounding each insert is shown in \fref{fig:PixelGeometry}, approximately equal in area to to each insert. This statistic aids in describing qualitative characteristics like clarity and visibility of the inserts, in spite of large quantitative errors in the low- and high-density regions (which will be discussed in section \ref{results}). 
\subsection{Optimal energy pairs}

Similar to the work of N{\"a}smark et al., \cite{Nasmark2021} an optimal high- and low-energy pair of VM images were selected to minimize the root-mean-square error (RMSE) across all inserts in the resulting SP images created from the $ \rho_e Z $ process. However, we noted that optimizing for minimal RMSE resulted in fairly large errors for all materials---even those similar to water, as in \fref{fig:ErrorBarA}. This is likely due to the extremely large errors in lung tissues or titanium, which can significantly affect the outlier-sensitive RMSE. Accordingly, the RMSE was also minimized using only groups of tissues (e.g. lung tissues, water-like tissues, bone tissues) and omitting all other inserts. Provided that all CT images are optimized with the same VM energy pair for a given tissue type and scanner, one could theoretically tailor the clinical reconstruction process based on the region of the body being imaged.
%
% -------------------------- Figure --------------------------------
\begin{figure}
\centering
\begin{subfigure}[t]{0.4\textwidth}
	\includegraphics[width=\textwidth]{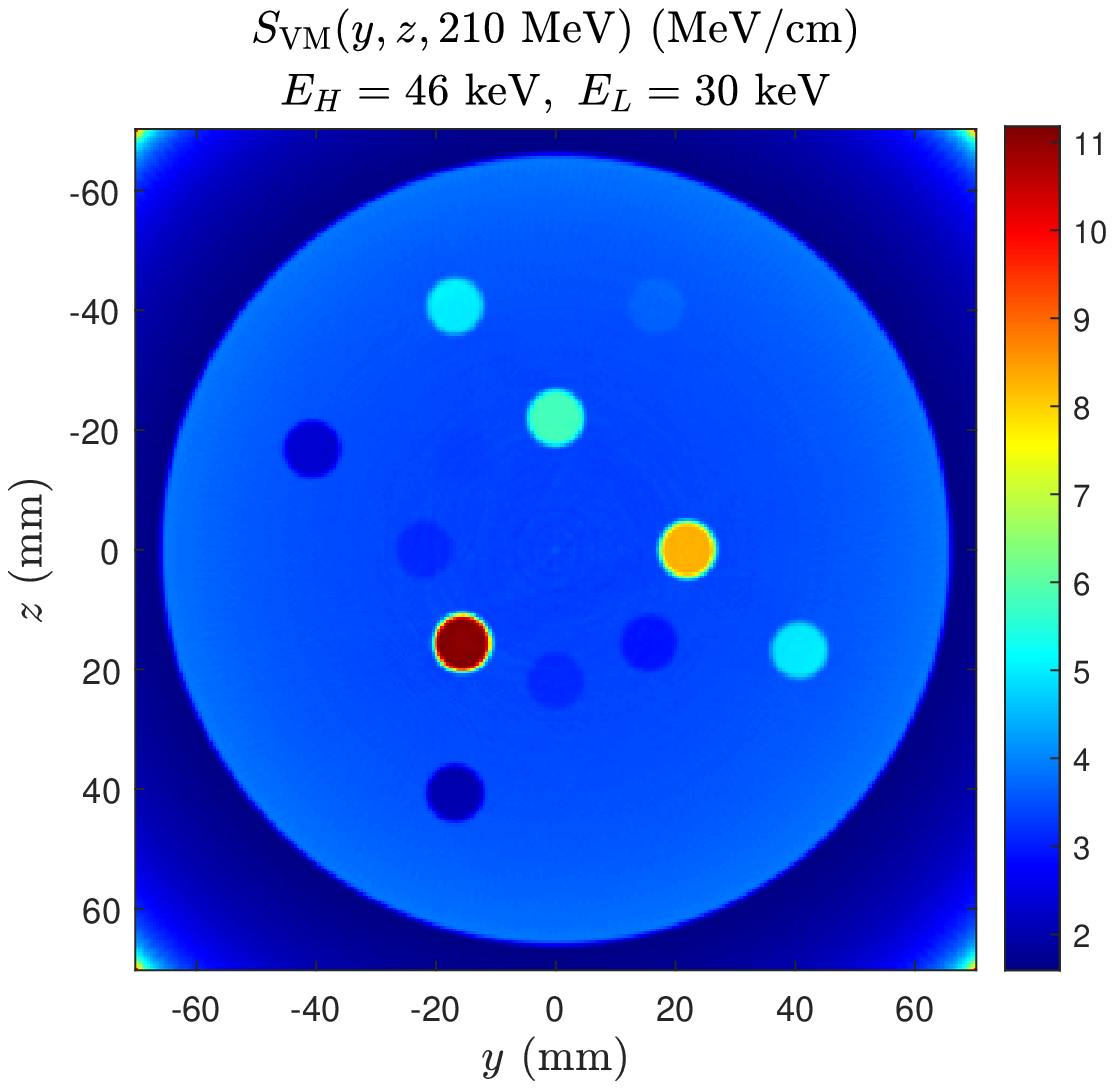}
	\subcaption{Standard configuration, optimized over all inserts.
	\label{fig:VM_ImagesA}}
\end{subfigure}
\hspace{3ex}
\begin{subfigure}[t]{0.39\textwidth}
	\includegraphics[width=\textwidth]{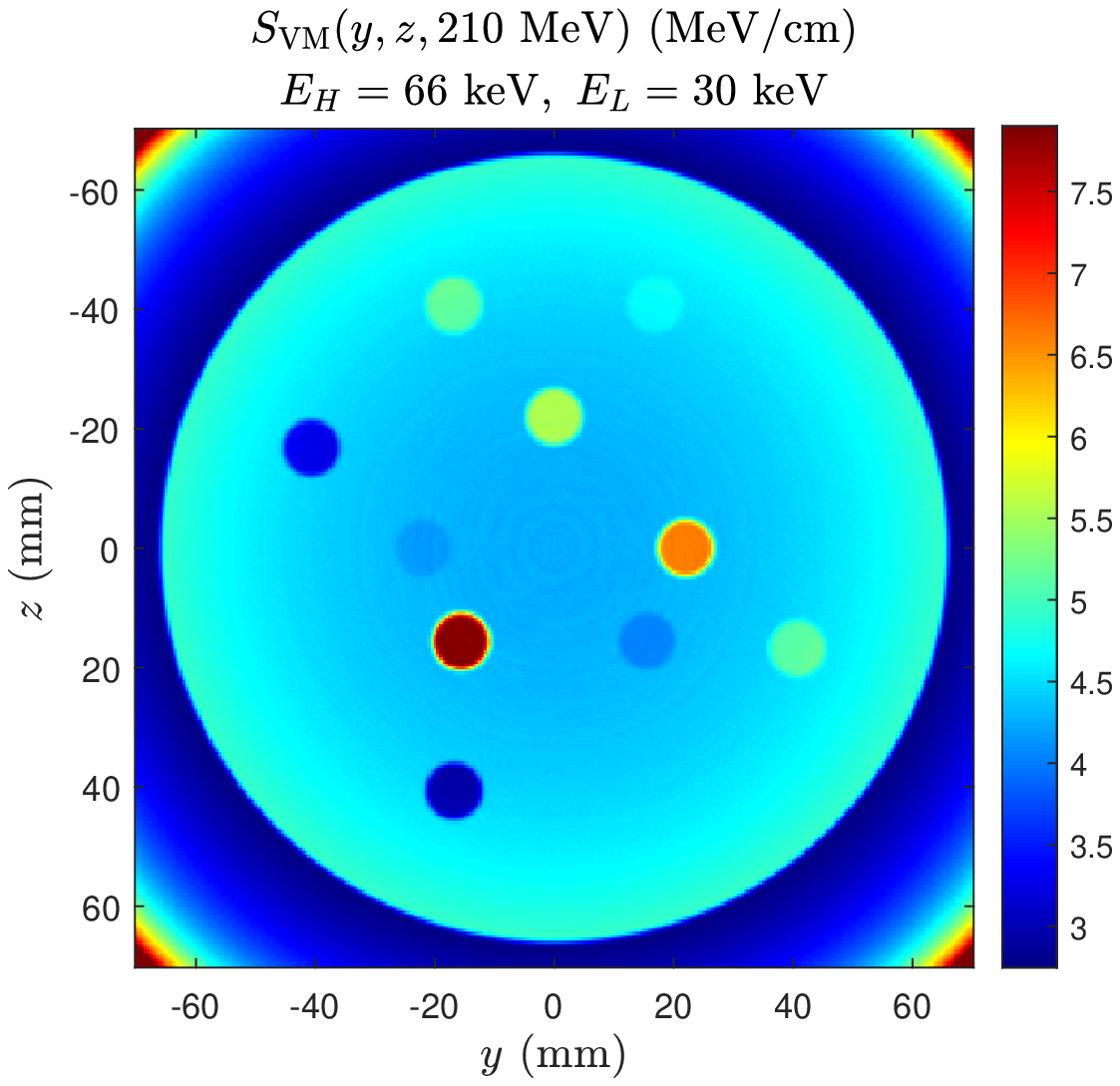}
		\subcaption{Standard configuration, soft-tissue optimized.
		\label{fig:VM_ImagesB}}
\end{subfigure}
\\[1ex]

\begin{subfigure}[t]{0.39\textwidth}
	\includegraphics[width=\textwidth]{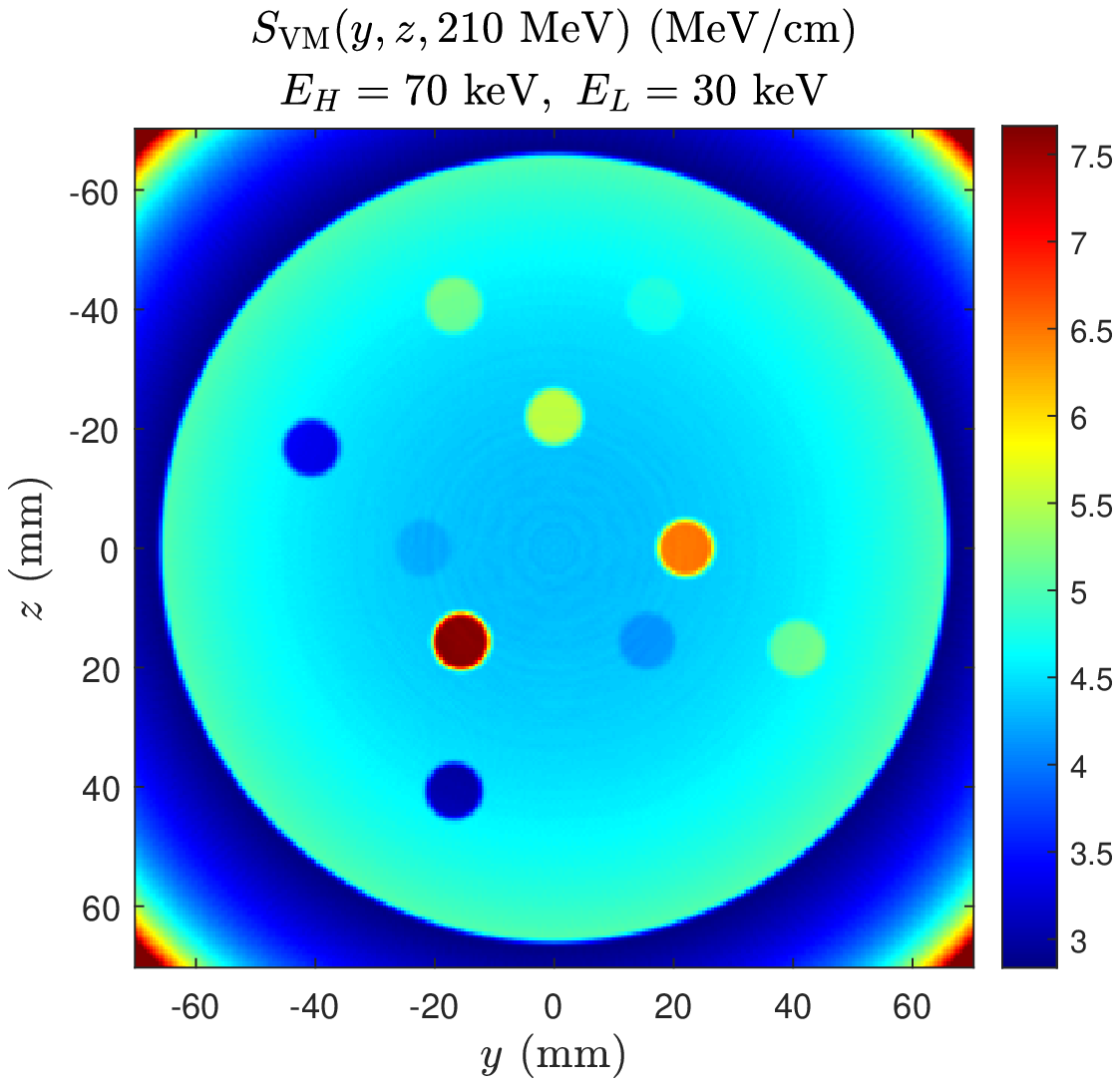}
		\subcaption{Standard configuration, lung tissue excluded.
		\label{fig:VM_ImagesC}}
\end{subfigure}
\hspace{3ex}
\begin{subfigure}[t]{0.39\textwidth}
	\includegraphics[width=\textwidth]{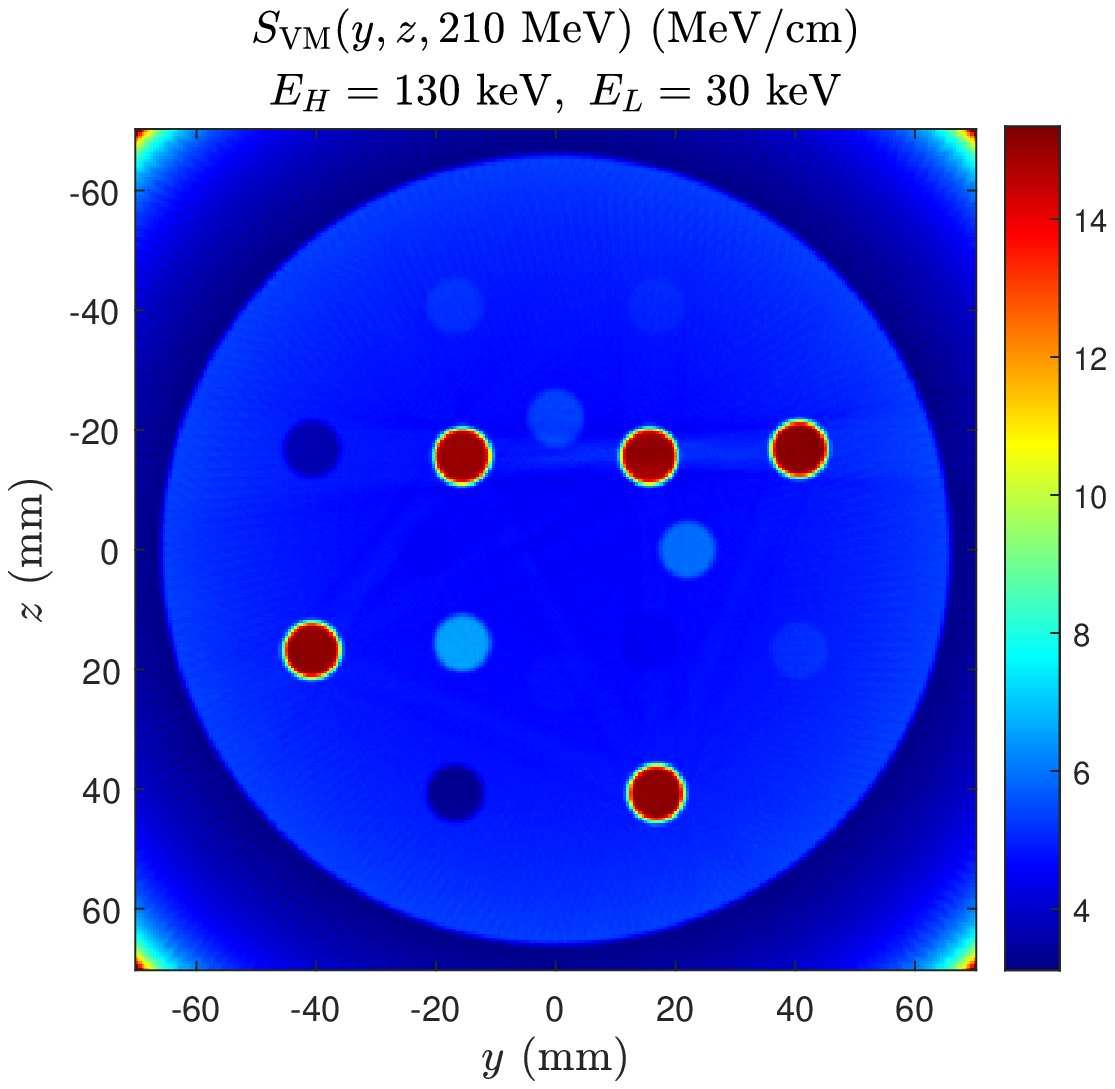}
		\subcaption{Titanium configuration, optimized over all inserts.
		\label{fig:VM_ImagesD}}
\end{subfigure}
\\[1ex]
\caption{SP images using the VM-SP-decomposition method, minimizing RMSE across specific tissue groups.
\label{fig:VM_Images}}
\end{figure}
% ------------------------------------------------------------------
%

To find the optimal energy pair for a tissue group, SPs were calculated using VM image-pairs for all energies from \SIrange{30}{130}{\keV}, with \SI{4}{\keV} spacing, such that $ E_H > E_L $. This range was chosen as the generally-accepted range of validity of the decomposition \eref{eq:MuDecomp}, \cite{Nasmark2021, JacksonHawkes1981} with the energy spacing chosen due to limits on computation time. The errors and contrast of optimal energy pairs for each scale parameter were compared.
%
% -------------------------- Figure --------------------------------
\begin{figure}[t]
\centering
\begin{subfigure}[t]{0.4\textwidth}
	\includegraphics[width=\textwidth]{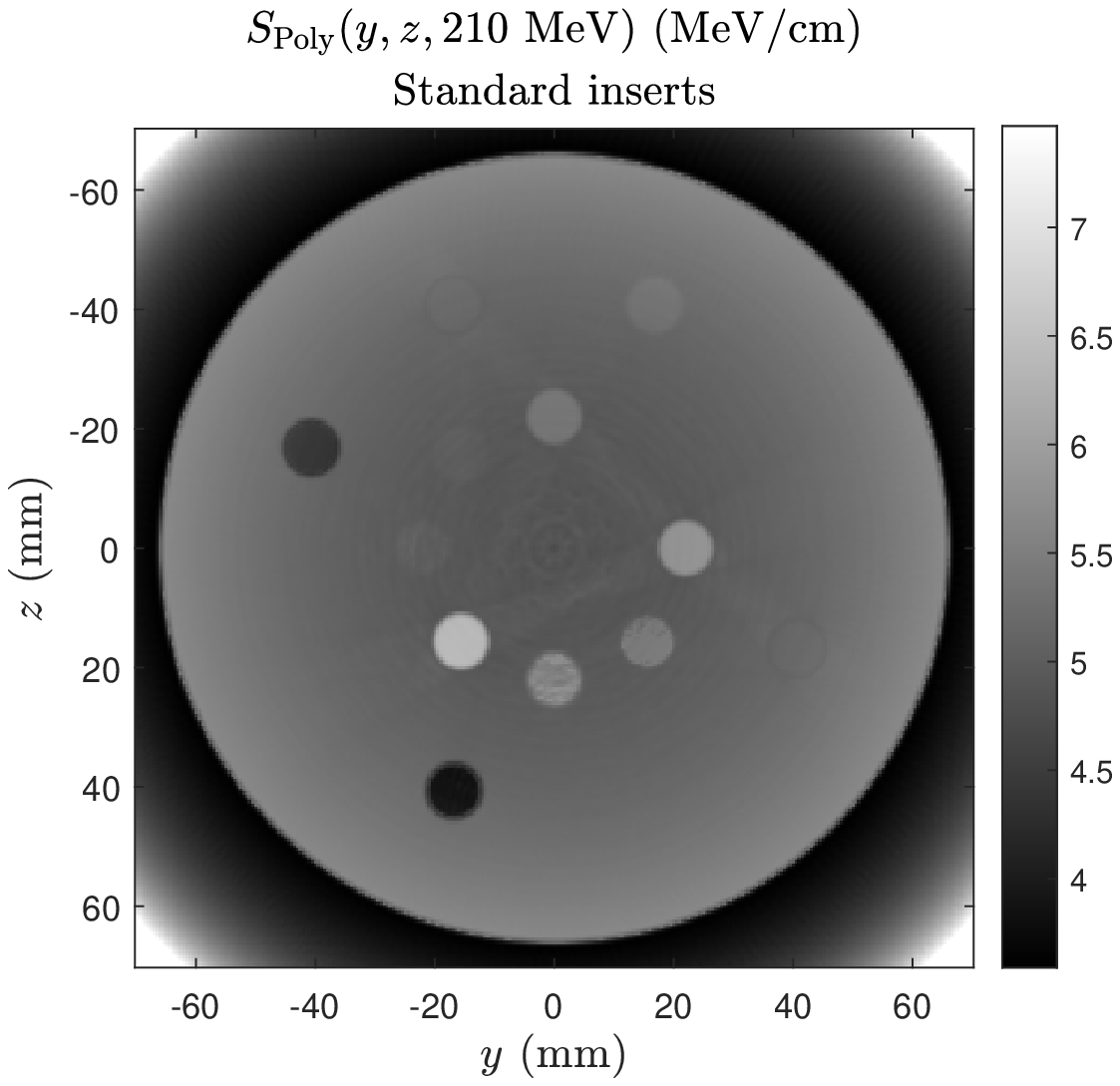}
		\subcaption{
		\label{fig:Poly_ImagesA}}
\end{subfigure}
\hspace{3ex}
\begin{subfigure}[t]{0.4\textwidth}
	\includegraphics[width=\textwidth]{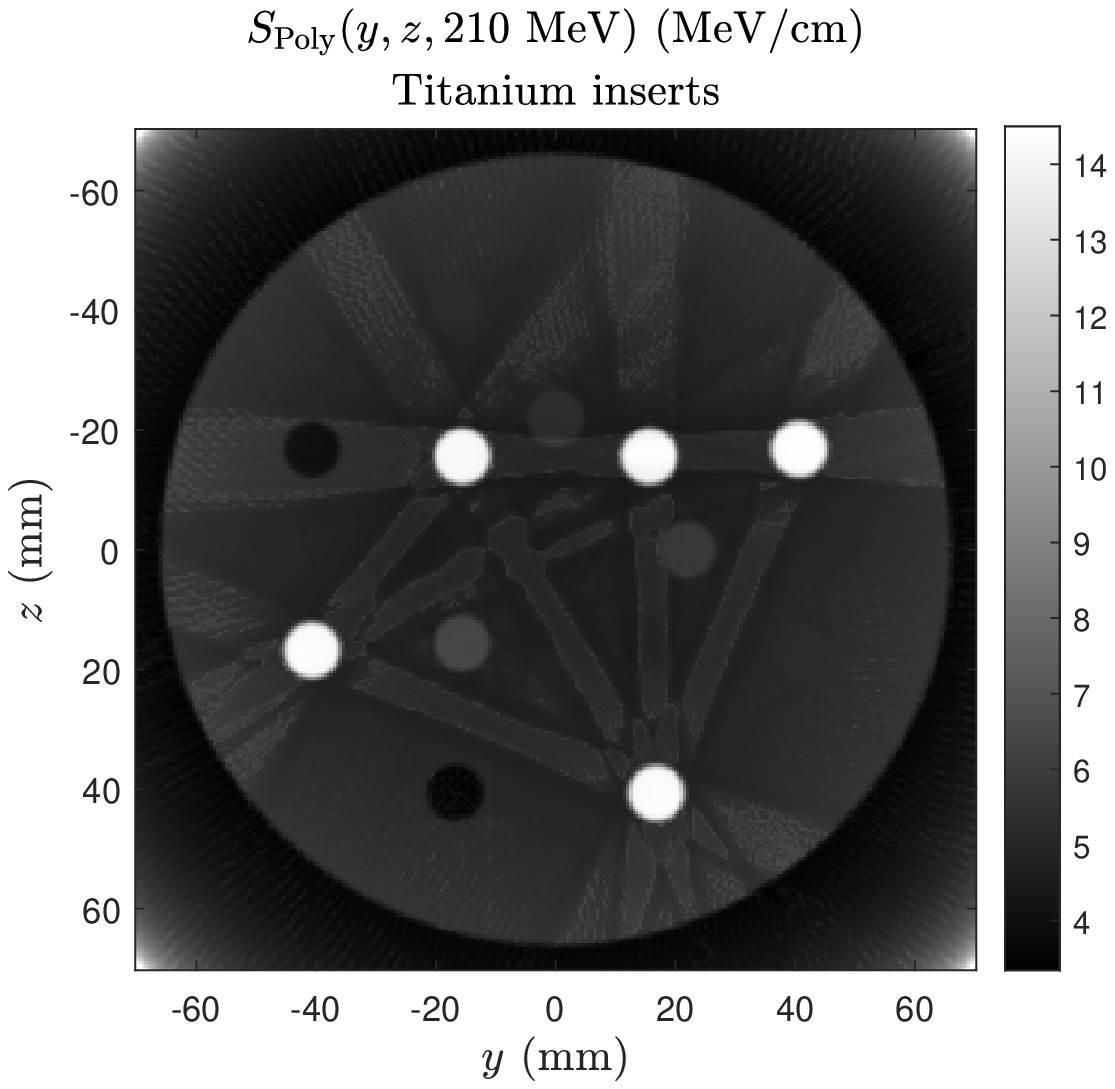}
		\subcaption{
		\label{fig:Poly_ImagesB}}
\end{subfigure}
\\[1ex]
\caption{Reconstructed SP images using the polychromatic CT-SP method. Note the presence of beam hardening streaks even in the standard configuration image.
\label{fig:Poly_Images}}
\end{figure}
% ------------------------------------------------------------------
%
\subsection{Comparison with a polychromatic CT-SP method}

In the interest of testing the validity of the VM-SP procedure against a pre-existing one, we have compared our results with an implementation of the polychromatic CT-SP method pioneered by Bazlova et al. \cite{Bazalova2008} We have used this polychromatic variant of the attenuation coefficient decomposition \eref{eq:MuDecomp}, as this method also does not require calibration scans, and is therefore convenient to implement with the same sets of Monte Carlo simulated CT data. See the appendix, or Yang et al., \cite{Yang2010} for a detailed description of this method. However, the $ I(\Zeff) $ curve of Bourque et al. is still utilized to convert EAN to MEE, rather than the log-linear fit of Yang et al. It should be mentioned that in our comparison, this polychromatic implementation is not utilized in conjunction with any x-ray pre-filtration or other forms of pre/post-processing to correct for beam hardening, and is therefore not a truly fair portrayal of the potential of the model presented by the aforementioned authors. This method is not claimed to be an appropriate benchmark, and is chosen solely as a representative of models that do not directly account for beam hardening and are also suitable for implementation using Monte Carlo data.
\section{Results} \label{results}
\subsection{Images and qualitative analysis}

Shown in \fref{fig:VM_Images} are reconstructed SP maps of standard and titanium configuration simulations, chosen with energy pairs that minimize RMSE for different tissue groups. While the quantitative accuracy of \fref{fig:VM_ImagesA}, \fref{fig:VM_ImagesB}, and \fref{fig:VM_ImagesC} are superior to \fref{fig:Poly_ImagesA} (see section \ref{Analysis}), brain tissue and water are nearly invisible in these images. It is also noteworthy that significant artifacting is seen in \fref{fig:Poly_ImagesB}. This is in contrast to the VM-SP images in \fref{fig:VM_Images}, which have no visible streaking. Also note the comparatively large streaks present in \fref{fig:Poly_ImagesA}, which contains no titanium. The contrast profiles for all images are similar (regardless of method used), but a noticeable increase in contrast is seen for most tissues using the VM-SP method---with the exception of the bone-optimizing energy pair in titanium configuration, which suffers from extremely high variances among titanium insert pixels.
%
% -------------------------- Figure --------------------------------
\begin{figure}
\centering
\begin{subfigure}[t]{0.46\hsize}
	\includegraphics[width=1\textwidth]{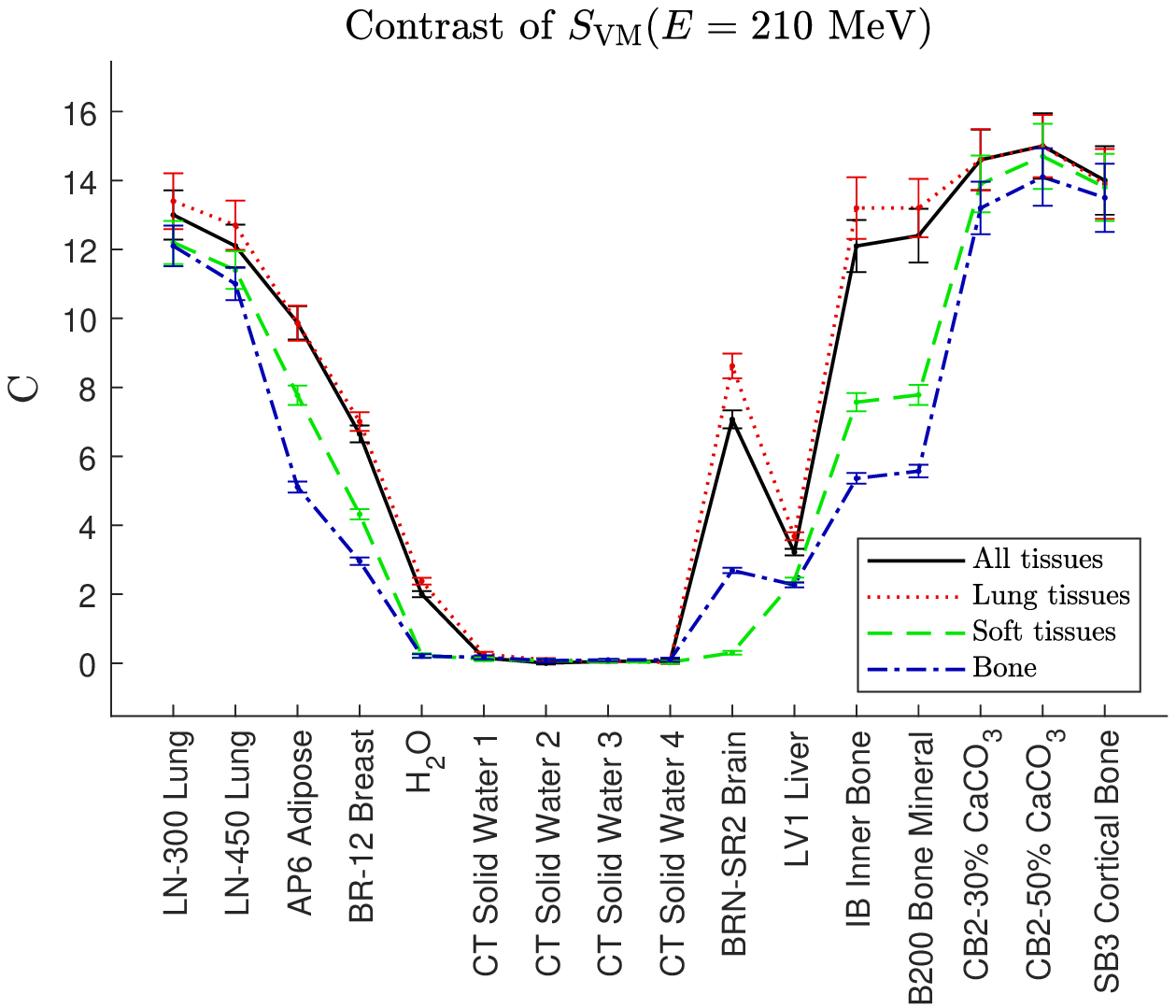}
		\subcaption{
		\label{fig:CNR_ImagesA}}
\end{subfigure}
\begin{subfigure}[t]{0.46\hsize}
	\includegraphics[width=1\textwidth]{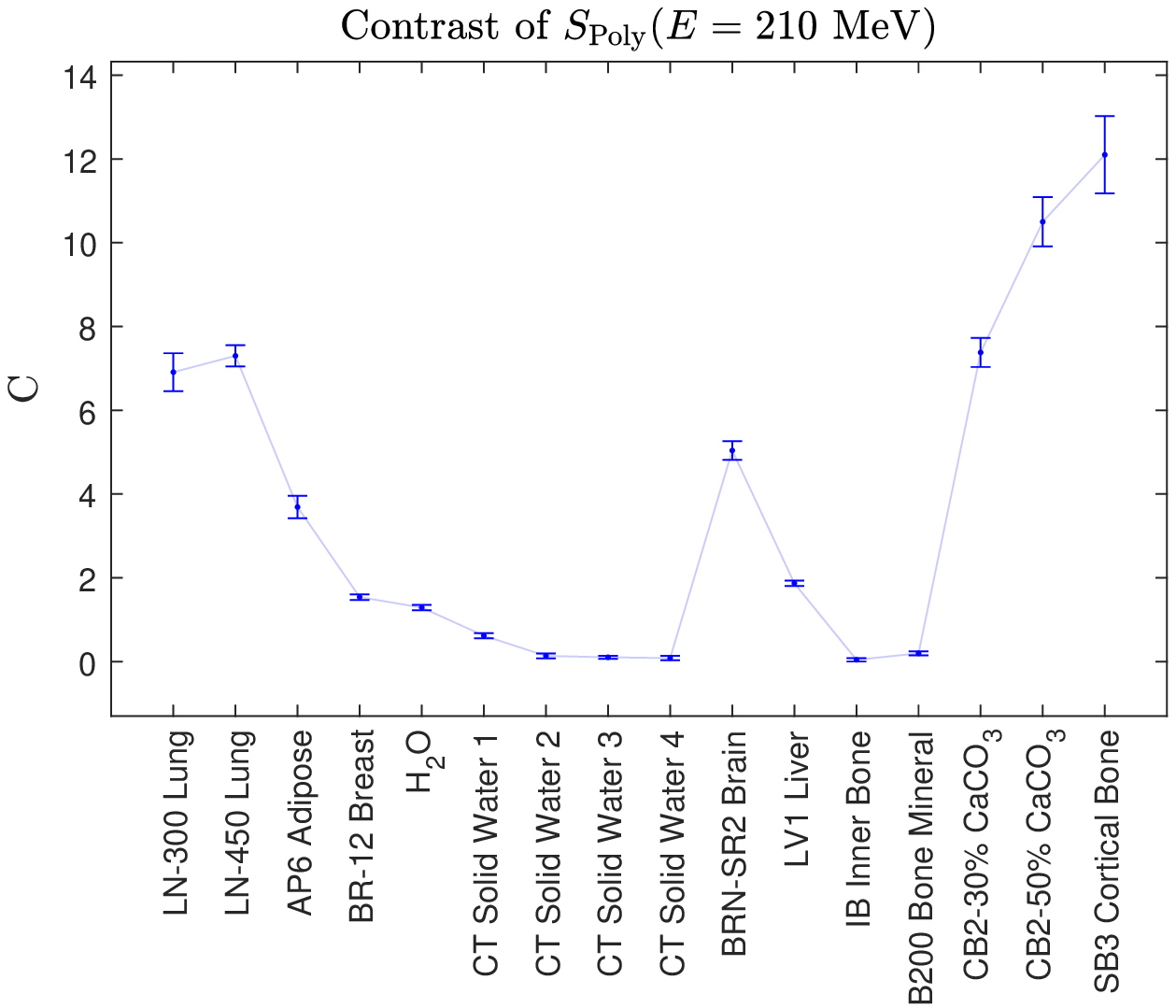}
		\subcaption{
		\label{fig:CNR_ImagesB}}
\end{subfigure}
\\[2ex]
	
\begin{subfigure}[t]{0.46\hsize}
	\includegraphics[width=1\textwidth]{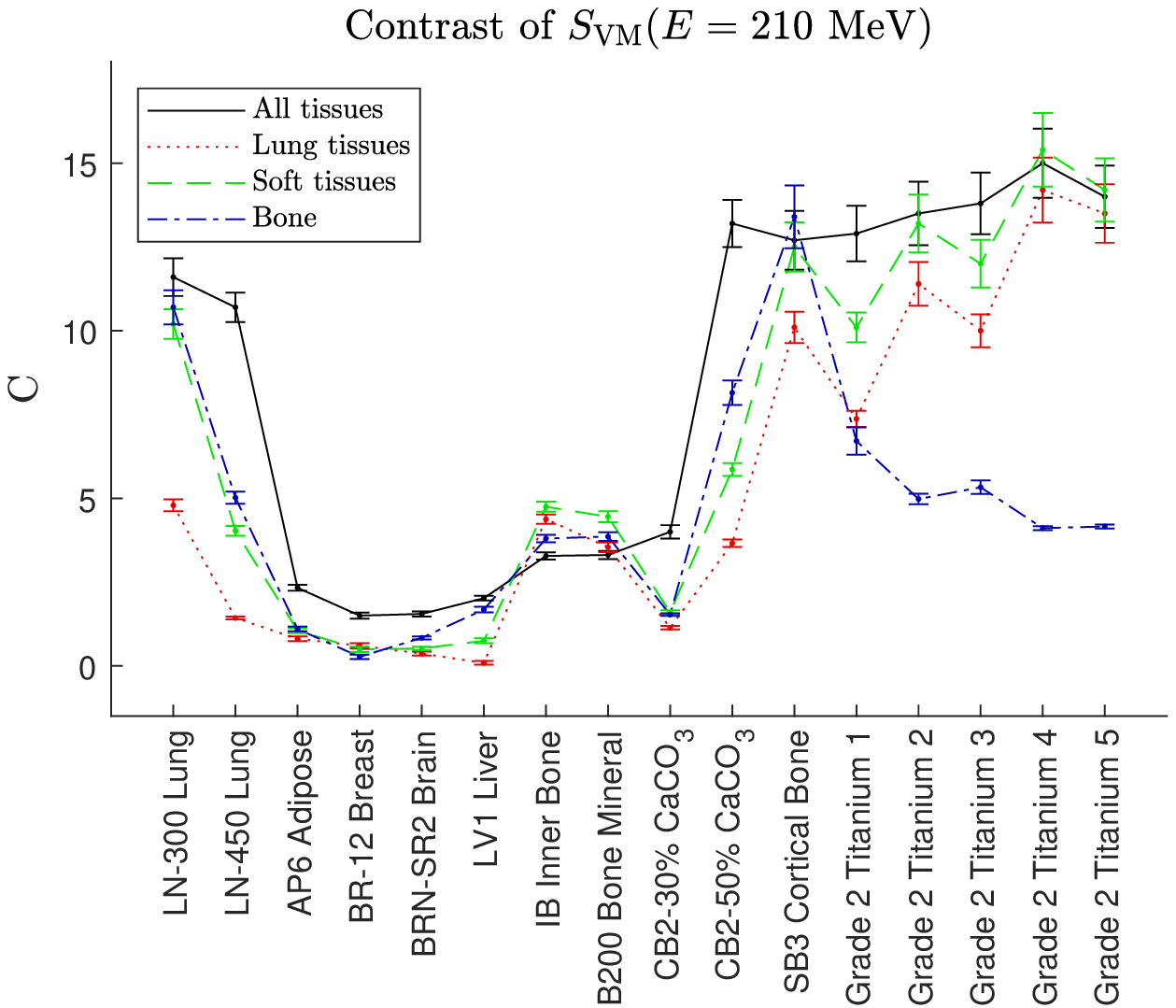}
		\subcaption{
		\label{fig:CNR_ImagesC}}
\end{subfigure}
\begin{subfigure}[t]{0.46\hsize}
	\includegraphics[width=1\textwidth]{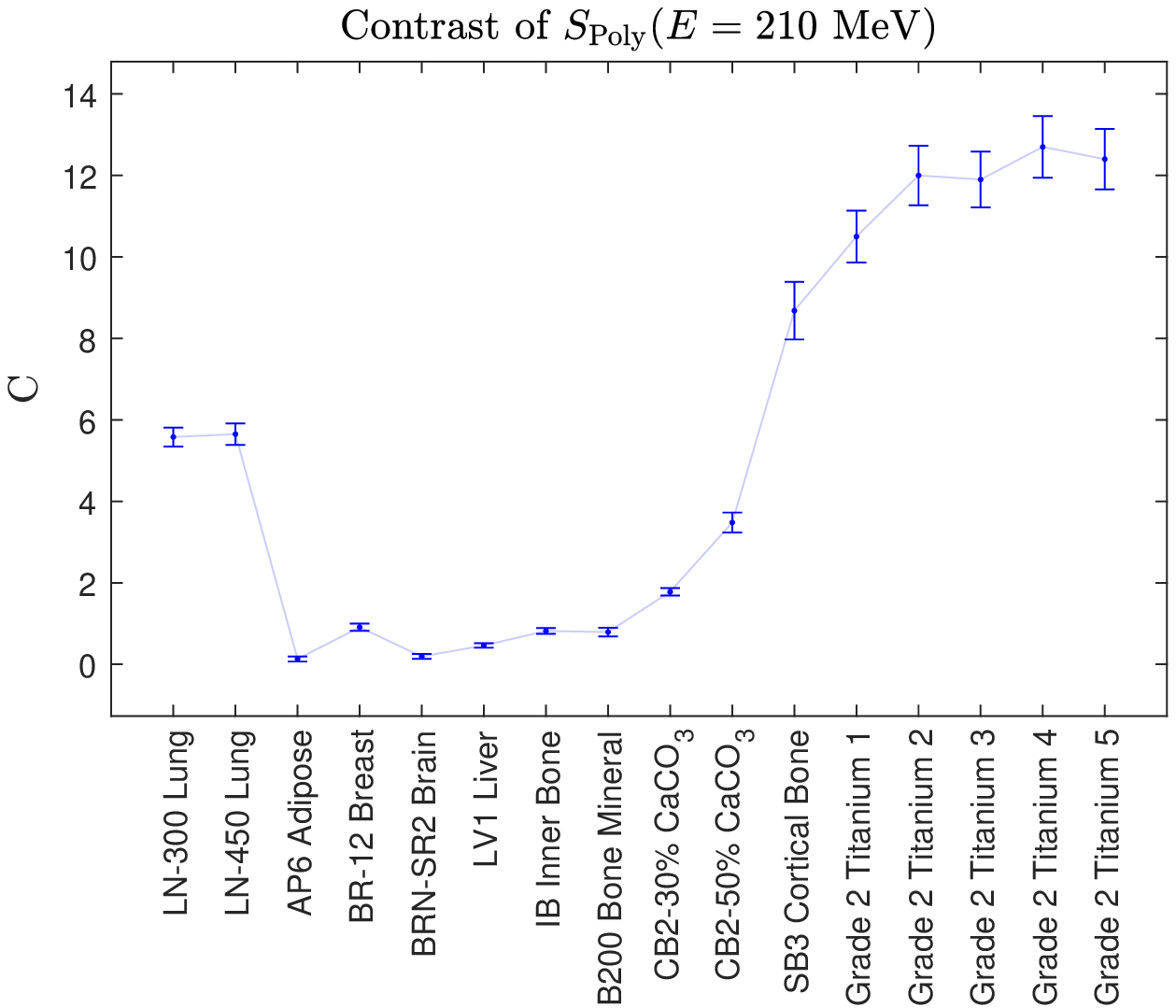}
		\subcaption{
		\label{fig:CNR_ImagesD}}
\end{subfigure}
\cprotect\caption{Contrast for images using VM energy pairs that minimize the RMSE for a given tissue group, as well as contrast for images using the polychromatic method. Error bars are one standard error, calculated with \num{10000} bootstrap samples of the insert/border regions using the Matlab \verb|bootstrp| function. \label{fig:CNR_Images}}
\end{figure}
% ------------------------------------------------------------------
%
\subsection{Error analysis} \label{Analysis}

\begin{figure}[tb]
\centering
\begin{subfigure}[t]{0.47\textwidth}
	\includegraphics[width=\textwidth]{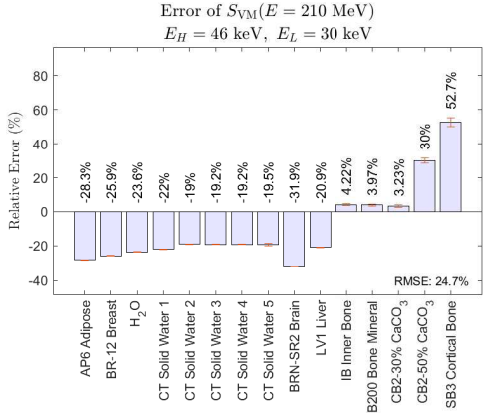}
		\subcaption{Standard configuration, optimized over all inserts.
		\label{fig:ErrorBarA}}
\end{subfigure}
\hspace{3ex}
\begin{subfigure}[t]{0.47\textwidth}
	\includegraphics[width=\textwidth]{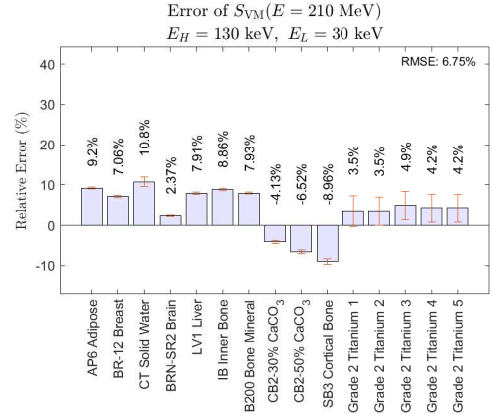}
		\subcaption{Titanium configuration, optimized over all inserts.
		\label{fig:ErrorBarB}}
\end{subfigure}
\\[1ex]

\begin{subfigure}[t]{0.47\textwidth}
	\includegraphics[width=\textwidth]{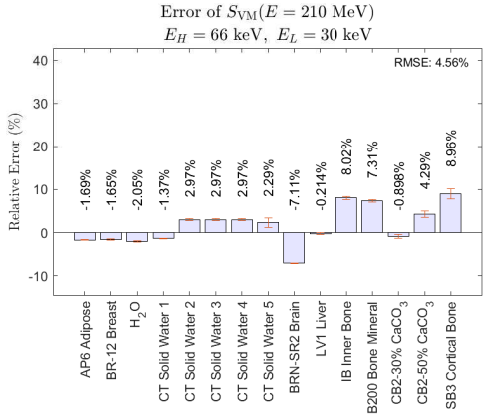}
		\subcaption{Standard configuration, soft-tissue optimizing.
		\label{fig:ErrorBarC}}
\end{subfigure}
\hspace{3ex}
\begin{subfigure}[t]{0.47\textwidth}
	\includegraphics[width=\textwidth]{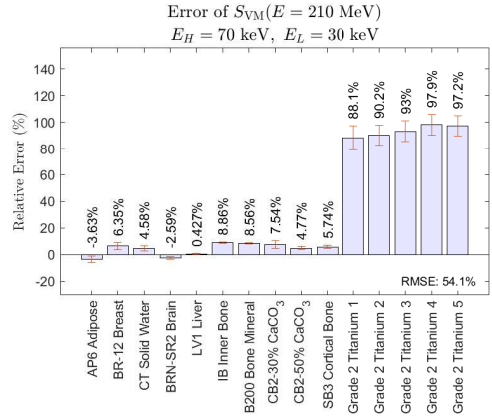}
		\subcaption{Titanium configuration, soft-tissue optimizing.
		\label{fig:ErrorBarD}}
\end{subfigure}
\\[1ex]
\caption{Relative SP errors for the phantom inserts, plotted in order of increasing mass density, for both insert configurations.
\label{fig:ErrorBars}}
\end{figure}

Plots of SP residual error for each insert are shown for the energy pairs that minimize the RMSE over subsets of inserts, given in \fref{fig:ErrorBars}. Errors in lung tissue are large outliers regardless of energy pair used, showing their ability to skew the RMSE: \fref{fig:ErrorBarA} and \fref{fig:ErrorBarD}---which include lung tissue in the RMSE calculation---show large errors on the order of \SIlist{20;8}{\percent} (respectively) for soft-tissues. Residual errors for lung tissues range from \SIlist{20;165}{\percent}, due to the small reference values of \SIlist{1.27;1.93}{\MeV\per\cm} at \SI{210}{\MeV}. Titanium inserts also have the potential to greatly affect the RMSE, as the titanium configuration image that minimizes only soft-tissue error contains extremely high errors.
%
% -------------------------- Figure --------------------------------
% ------------------------------------------------------------------
%

\begin{figure}[t]
\centering
\begin{subfigure}[t]{0.44\textwidth}
	\includegraphics[width=\textwidth]{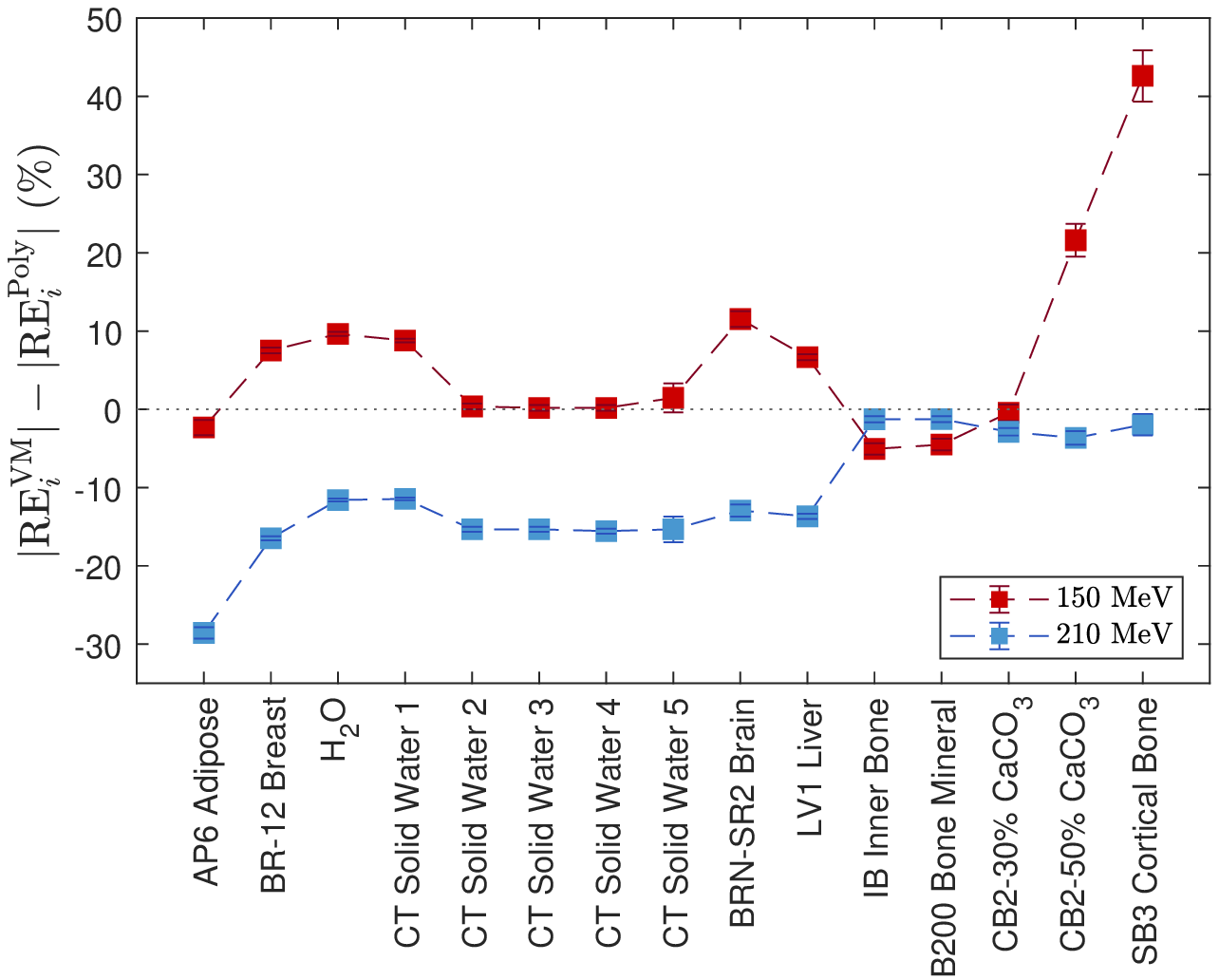}
		\subcaption{
		\label{fig:PolyCompare_St}}
\end{subfigure}
\hspace{3ex}
\begin{subfigure}[t]{0.44\textwidth}
	\includegraphics[width=\textwidth]{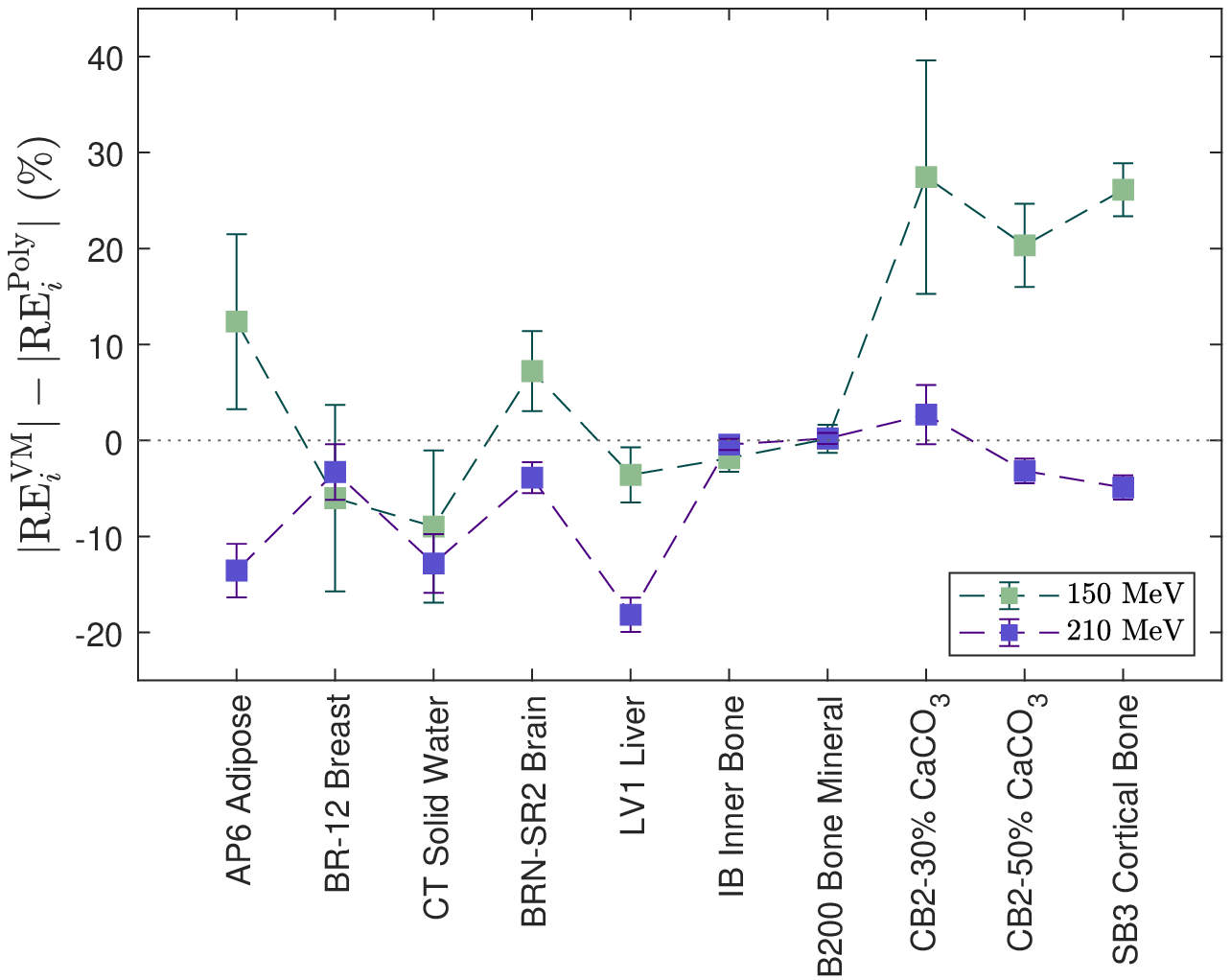}
		\subcaption{
		\label{fig:PolyCompare_Ti}}
\end{subfigure}
\\[1ex]
\caption{Differences in residual error between soft-tissue optimized VM-SP images and polychromatic SP images. Error bars are one standard error of the mean, adding the errors of \fref{fig:ErrorBars} in quadrature. \label{fig:PolyCompare}}
\end{figure}

Compared to the VM method, the polychromatic method (\fref{fig:PolyCompare}) exhibits errors larger than \SI{10}{\percent} across almost all inserts---including water-equivalent tissues. However, the polychromatic method demonstrates significantly lower titanium error in general. \Fref{fig:ErrorBarD} and \fref{fig:ErrorBarB} show that soft-tissues and high-density tissues can not be simultaneously optimized using the parameters chosen for this work. This ill-behaved behavior of the titanium configuration may even indicate that the presence of titanium extends the model beyond its range of validity. Similarly, the large lung tissue errors may in part be attributable to low-density tissues being unsuited to the VM-SP method. For comparison, the polychromatic CT-SP method is compared to the soft-tissue optimizing energy pairs of the VM-SP method in figure \fref{fig:PolyCompare}. To test both models at different ends of the therapeutic proton energy range, the models are also compared at \SI{150}{\MeV}, where the failure of the VM-SP method at high densities becomes even more pronounced.
%
% -------------------------- Figure --------------------------------
% ------------------------------------------------------------------
%

\begin{figure}[t]
\centering
\begin{subfigure}[t]{0.46\textwidth}
	\includegraphics[width=\textwidth]{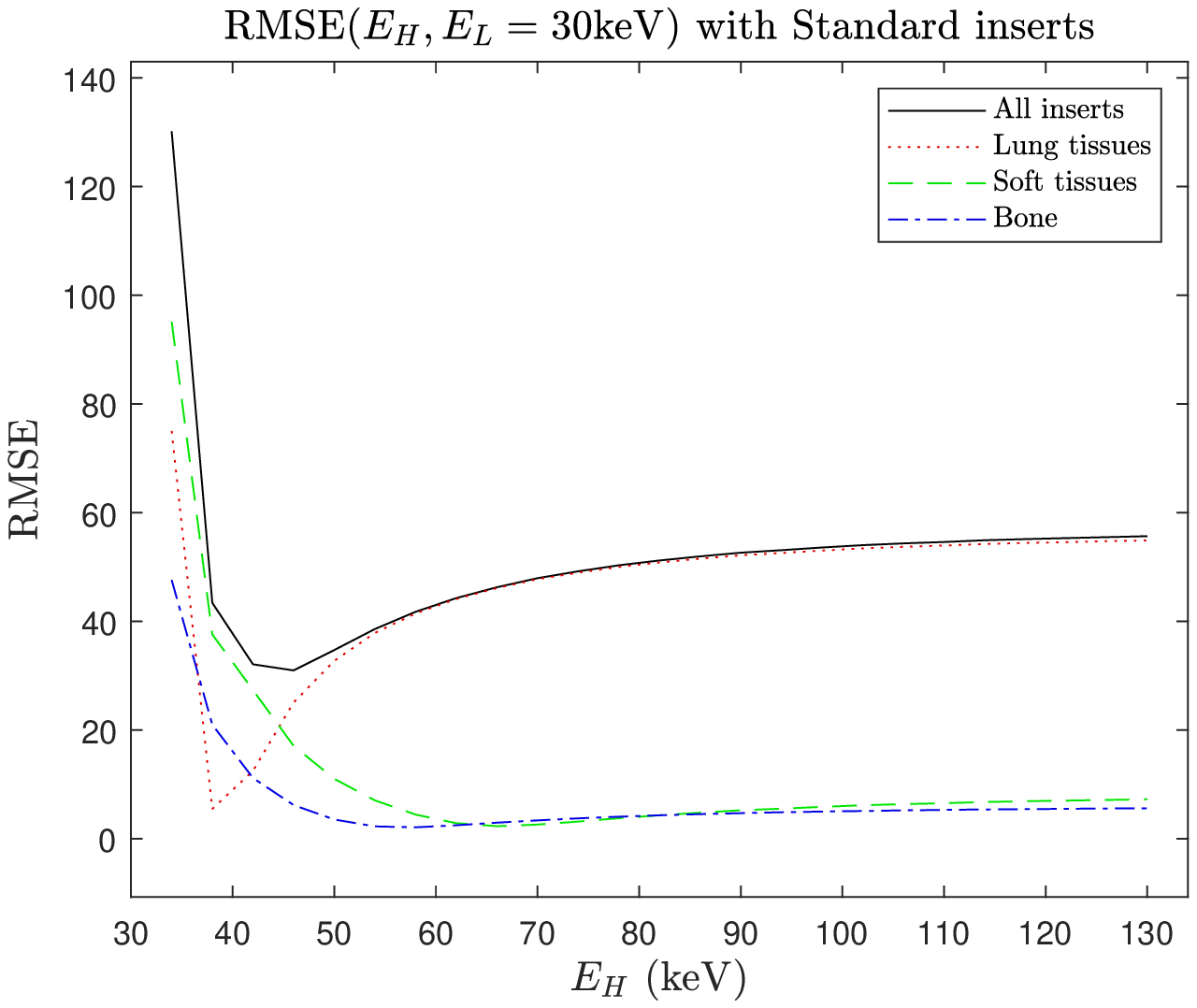}
		\subcaption{
		\label{fig:minLinesE}}
\end{subfigure}
\hspace{3ex}
\begin{subfigure}[t]{0.46\textwidth}
	\includegraphics[width=\textwidth]{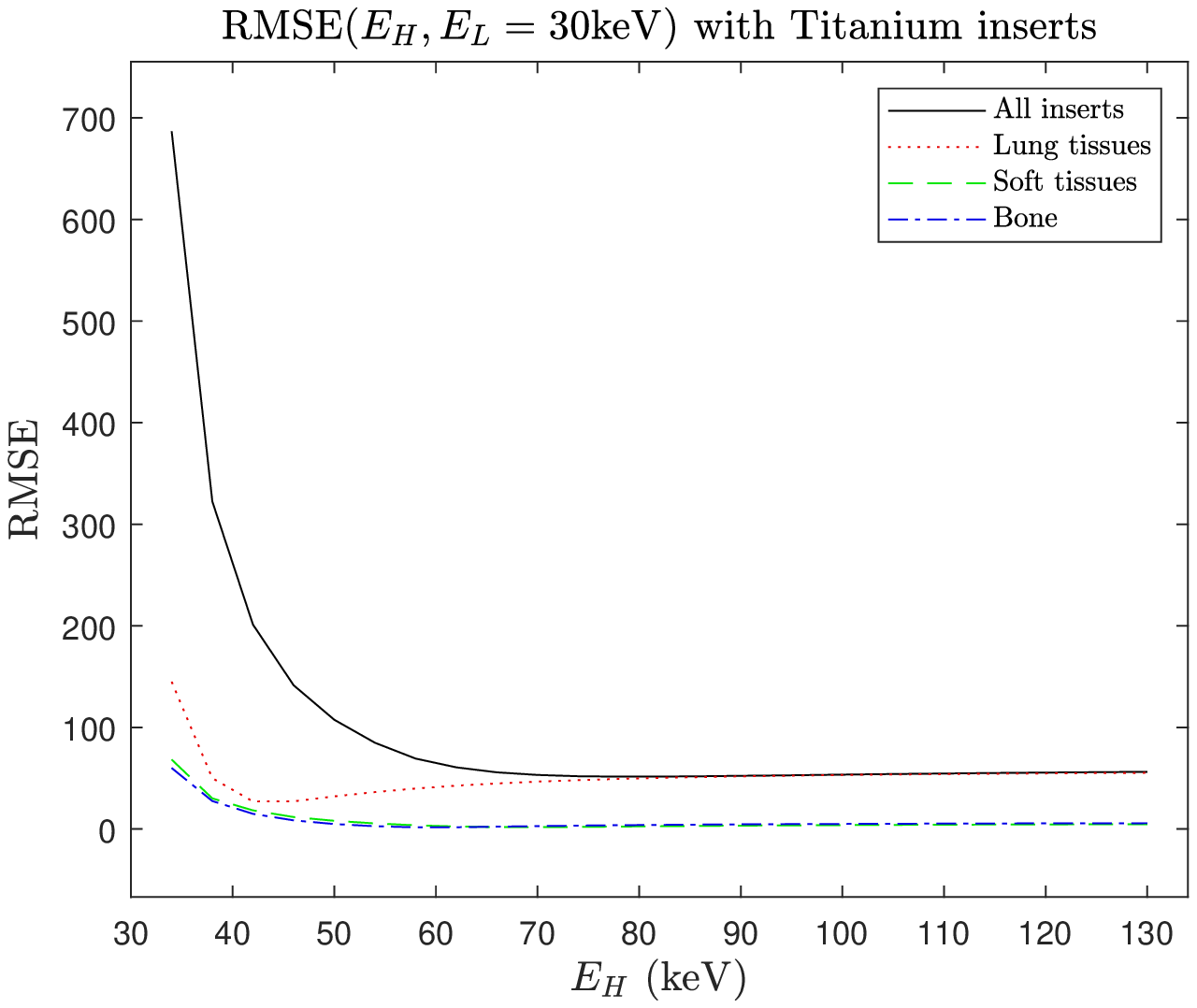}
		\subcaption{
		\label{fig:minLinesF}}
\end{subfigure}
\\[1ex]
\caption{RMSE, as a function of $ E_H $ with $ E_L = \SI{30}{\keV} $,  with both standard and titanium inserts. The poor performance of the  RMSE optimized over all inserts is seen, in response to the presence of lung tissue.
\label{fig:minLines}}
\end{figure}

\begin{figure}
\centering
\includegraphics[width=0.7\textwidth]{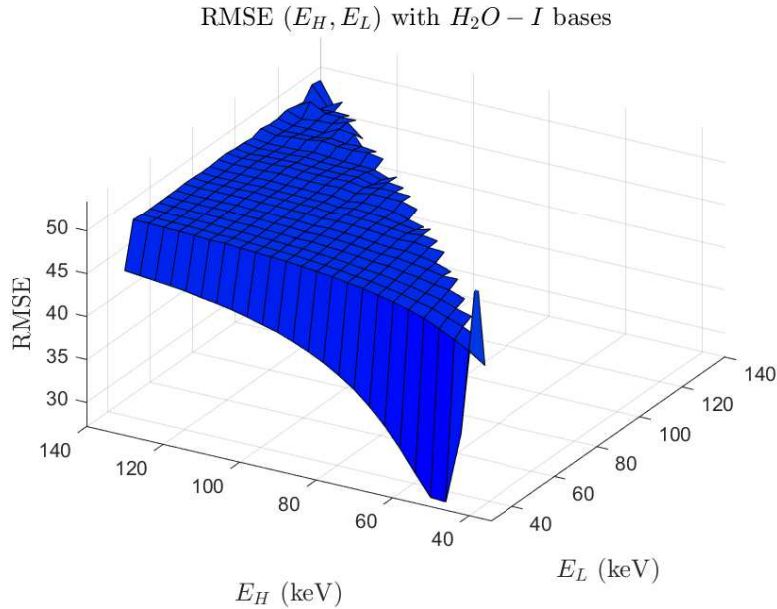}
\caption{RMSE as a function of $ E_H $ and $ E_L $ for the standard configuration phantom, optimized over all tissues.  \label{fig:RMSE_Sheet}}
\end{figure}

For use in a clinical setting, it is desirable to find a region of the 2D $ (E_H, E_L) $ energy space that optimizes a summary statistic for images with specific kinds of tissues constituting the majority of the image (e.g. an energy pair that optimizes images that contain mostly lung tissue, or images with titanium implants, etc). To examine this, the RMSE has been calculated using only tissues from one tissue type at a time: only lung tissues, only soft tissues, or only bone tissues. The optimal energy pairs for both configurations and all three tissue optimizations involve $ E_L \approx \SI{30}{\keV} $, so the optimal parameters are plotted as a function of $ E_H $ with constant $ E_L $ in \fref{fig:minLines}. For the standard configuration phantom, a minimum is achieved between approximately \SIlist{40;70}{\keV} (depending on the optimized tissue subset), after which the RMSE gradually approaches a constant value. For the titanium configuration, there does not appear to be a local minimum, and higher values of $ E_H $ yield lower RMSE. Note that as $ E_H $ increases, the RMSE calculated over all inserts converges with the RMSE calculated using only lung tissue, demonstrating how deeply the parameter is affected by outlier tissues with large errors. In general, however, the RMSE is largely unaffected by the choice of $ E_H $, provided that no extreme outlier tissues are present, $ E_H > \SI{60}{\keV} $, and $ E_L $ remains as small as possible (as can be seen in \fref{fig:RMSE_Sheet}). The lack of local extreme in these plots may indicate a failure of the model with the chosen basis materials and parameters.
\section{Discussion}

While the method presented here is an appreciable confirmation of concept, it has faults and shortcomings that must be addressed. Firstly, the optimal energies almost always utilize $ E_L < \SI{30}{\keV} $ if allowed, which is typically considered below the breakdown point of \eref{eq:MuDecomp}, where photoelectric and incoherent cross sections are similar in magnitude. This may be the result of an ill-suited $ \rho_e Z $-decomposition equation (as in \eref{eq:MuDecomp}). Results may be improved through a change of photoelectric basis in \eqref{eq:Alvarez} or a change in the power law of $ Z^n $ in \eref{eq:MuDecomp}. Iodine performed better than other photoelectric bases tested in this work (titanium, calcium, and aluminum), but the presence of a K-edge in the diagnostic energy range can introduce errors, and there may be a more suitable photoelectric basis for other phantoms and setups. Jackson and Hawkes \cite{JacksonHawkes1981} note that the power of the $ Z^n $-dependence in \eref{eq:MuDecomp} is variable depending on the data set it is modeled with, as coherent scattering overtakes the photoelectric effect in likelihood beyond about \SI{70}{\keV} for water-like materials. Changing both $ n $ and the attenuation bases may produce better results.

In additional to faults in the proposed model, simplifications have been made in the acquisition of Monte Carlo data, which need to be addressed for use in a clinical setting. While it is true that an x-ray tube's spectrum can be generally predicted provided the tube's anode angle, anode material, and inherent amount material-equivalent filtration are known, \cite{Boone1988, Tucker1990} any given tube will deviate from theoretical models, and measurements of a clinical scanner's spectrum is necessary to utilize the VM-SP model presented here. Further simulations are also required to test the model under more realistic conditions, such as: a curved detector plane in conjunction with an x-ray fan-beam, and tallying that simulates the imperfect response of a real detector. However, the choice to model a straight-line detector may be justifiable, as fan-beam data can be re-sorted into parallel-projection data. 

Most importantly, it must be stated that while the proposed VM method reduces the total error and removes more beam-hardening streaks than our implementation of the polychromatic method, the magnitude of the errors are still high, compared to those reported in the literature for other spectra-dependent, DECT-based $ \rho_e Z $-decomposition models, \cite{Bazalova2008, Yang2010, VanAbbema2015, Dedes2019, Tanaka2020, Nasmark2021} which all report errors of less than or around \SI{1}{\percent} for most tissues with densities in the vicinity of \SI{1}{\g\per\cm\cubed}. This may be a consequence of the unaltered FBP algorithm used for reconstruction of the attenuation images, as similar errors for lung, bone, and titanium are also present for both raw DECT and traditional x-ray attenuation images used to generate stopping power images. Consequently, this method could be improved with a more accurate x-ray CT reconstruction, such as an iterative algebraic algorithm.

%
% -------------------------- Figure --------------------------------
% ------------------------------------------------------------------
%
%
% -------------------------- Figure --------------------------------
% ------------------------------------------------------------------
%
\section{Conclusion}

A recently proposed, dual-energy-based stopping power determination method has been slightly modified and implemented, using Monte Carlo simulated x-ray CT data. This model utilizes pairs of simulated CT scans to create proton stopping power images in a way that maximizes the number of tissues with low residual error, utilizing the potential of DECT imaging to reduce beam-hardening artifacts and eliminate the inherent degeneracy of the CT stoichiometric calibration. As this model is dependent only on spectral information from the x-ray tube, it can be applied to any scanner capable of producing low- and high-energy spectra, as long as these spectra are known. To simulate realistic CT data, we have modeled a widely-used tissue calibration phantom and based the incident spectra on the parameters of a currently-available dual-energy CT scanner. Comparing this method to another dual-energy $ \rho_e Z $ decomposition method, we see a slight reduction in error for all non-lung tissues, as well as a noticeable decrease in beam hardening artifacting in images with many high-density materials. Consequently, our results demonstrate the validity of this model, lending evidence to the claim that it may have the potential to improve the $ \rho_e Z $-based SP determination process.
\section*{Appendix}

%\addcontentsline{toc}{section}{\numberline{}Appendix}
The polychromatic $ \rho_e Z $-decomposition method described by Bazalova et al. \cite{Bazalova2008} is briefly given here for completeness, describing the comparison method used in this study. Much like \eref{eq:MuDecomp}, this method decomposes the x-ray attenuation into photoelectric and scattering components as a function of $ \Zeff $ and $ \rho_e $, but assuming a polychromatic x-ray source. The spectrally-averaged attenuation is given by
%
% -------------------------- Equation ------------------------------
\begin{eqnarray}
\mueff_k = \rho_e \sum_i \S_k(E_i) \Big( \Zeff^4 F(\Zeff, E_i) + G(\Zeff, E_i) \Big) \Delta E_i,
\label{eq:effMu}
\end{eqnarray}
% ------------------------------------------------------------------
%
which is iteratively solved for $ \Zeff $, like \eref{eq:Zeff}, using the following equation given by Yang et al. \cite{Yang2010}:
%
% -------------------------- Equation ------------------------------
\begin{eqnarray}
\Zeff^4 - \frac{ \mueff_H \times \Sigma_{L}^{G}(\Zeff) - \mueff_L \times \Sigma_{H}^{G}(\Zeff) }{ \mueff_L \times \Sigma_{H}^{F}(\Zeff) - \mueff_H \times \Sigma_{L}^{F}(\Zeff) } = 0,
\label{eq:polyZeff}
\end{eqnarray}
% ------------------------------------------------------------------
%
with the following additional factors defined for ease of reading:
%
% -------------------------- Equation ------------------------------
\numparts
\begin{eqnarray}
& \Sigma_{k}^{G}(\Zeff) = \sum_{i} \S_k(E_i) G(\Zeff,E_i) \Delta E_i, \label{eq:sumG} \\
& \Sigma_{k}^{F}(\Zeff) = \sum_{i} \S_k(E_i) F(\Zeff,E_i) \Delta E_i. \label{eq:sumF}
\end{eqnarray}
\endnumparts
% ------------------------------------------------------------------
%
ED is then solved by inserting \eref{eq:polyZeff} into the following equation, in analogy with \eref{eq:rhoe}:
%
% -------------------------- Equation ------------------------------
\begin{eqnarray}
\rho_e = \frac{ \mueff_L \times \Sigma_{H}^{F}(\Zeff) - \mueff_H \times \Sigma_{L}^{F}(\Zeff) }{ \Sigma_{H}^{F}(\Zeff) \times \Sigma_{L}^{G}(\Zeff) - \Sigma_{L}^{F}(\Zeff) \times \Sigma_{H}^{G}(\Zeff) }. \label{eq:polyED}
\end{eqnarray}
% ------------------------------------------------------------------
%
Rather than utilizing VM x-ray attenuation images, this method uses traditional, effective-energy (i.e. spectrally-averaged) attenuations $ \mueff_k $ as input. Note the use of the tube entrance spectrum in \eref{eq:effMu}. In reality, the tube spectra $ \S_k $ harden as they traverse the phantom, creating a significantly different exit spectrum along every unique x-ray path in the set of CT projections. By assuming that $ \S_k = \S_{k, \mathrm{exit}} $, beam hardening is neglected theoretically.

\section*{References}
\bibliography{ms}
\bibliographystyle{unsrt.bst}

\end{document}